\documentclass[acmsmall]{acmart}
\usepackage{algpseudocode}
\usepackage{subcaption}
\usepackage{algorithm}
\usepackage{multirow}
\usepackage{enumitem}
\usepackage{bm}
\usepackage{xspace}
\usepackage{marvosym}

\usepackage{bbding}
\AtBeginDocument{%
  }

\newcommand{\ie}{\emph{i.e.,}\xspace}
\newcommand{\eg}{\emph{e.g.,}\xspace}

\newcommand{\paratitle}[1]{\vspace{1.5ex}\noindent\textbf{#1}}

\newcommand{\ignore}[1]{}

\begin{document}

\title[Dual-Stream MLP is All You Need for CTR Prediction]{\texorpdfstring{Dual-Stream MLP is All You Need for CTR Prediction}{Dual-Stream MLP is All you need for CTR predictions}}

\author{Kesha Ou}
\orcid{0009-0006-8557-5437}
\affiliation{
  \institution{Gaoling School of Artificial Intelligence, Renmin University of China}
  \city{Beijing}
  \country{China}
}
\authornote{Both authors contributed equally to this research.}
\authornote{Also with Beijing Key Laboratory of Research on Large Models and Intelligent Governance.}
\email{keishaou@gmail.com}

\author{Zhen Tian}
\orcid{0000-0001-5569-2591}
\affiliation{
    \institution{ByteDance}
    \city{Beijing}
    \country{China}
}
\authornotemark[1]
\email{chenyuwuxinn@gmail.com}

\author{Wayne Xin Zhao$^\spadesuit$}
\orcid{0000-0002-8333-6196}
\affiliation{
    \institution{
    Gaoling School of Artificial Intelligence,  
    Renmin University of China}
    \city{Beijing}
    \country{China}
}
\authornotemark[2]
\email{batmanfly@gmail.com}
\thanks{$\spadesuit$ Corresponding author.}

\author{Long Zhang}
\orcid{0009-0003-4769-5030}
\affiliation{
    \institution{Meituan}
    \city{Beijing}
    \country{China}
}
\email{zhanglong40@meituan.com}

\author{Sheng Chen}
\orcid{}
\affiliation{
    \institution{Meituan}
    \city{Beijing}
    \country{China}
}
\email{chensheng19@meituan.com}

\author{Ji-Rong Wen}
\orcid{0000-0002-9777-9676}
\affiliation{
    \institution{
    Gaoling School of Artificial Intelligence,  
    Renmin University of China}
    \city{Beijing}
    \country{China}
}
\authornotemark[2]
\email{jrwen@ruc.edu.cn}

\thanks{This work was supported by the National Natural Science Foundation of China No.~92470205 and Beijing Major Science and Technology Project No.~Z251100008425002.}

\newcommand{\tba}{\textcolor{red}{xxx }}
\newcommand{\outd}{\textcolor{red}{[Outdated]}~}
\newcommand{\tabincell}[2]{\begin{tabular}{@{}#1@{}}#2\end{tabular}}

\definecolor{dark2green}{rgb}{0.1, 0.65, 0.3}
\definecolor{dark2orange}{rgb}{0.9, 0.4, 0.}
\definecolor{dark2purple}{rgb}{0.4, 0.4, 0.8}
\newcommand{\first}[1]{\textbf{#1}}
\newcommand{\second}[1]{\underline{#1}}
\newcommand{\third}[1]{\textbf{\textcolor{dark2purple}{#1}}}

\begin{abstract}
Click-through rate~(CTR) prediction holds a pivotal role in online advertising and recommendation systems, where even small improvements can significantly boost revenue. 
Existing research primarily focuses on designing dual-stream architectures to capture effective  complex feature interactions from both explicit and implicit perspectives. 
However, these approaches are faced with two major challenges: 1) the high complexity of feature interaction learning, which increases computational demands and the overfitting risk, and 2) the imbalance between explicit and implicit modules, where one module's output may dominate the final prediction. 
To address these issues, in this paper, we propose \textbf{D}ual-\textbf{S}tream \textbf{MLP}~(\textbf{DS-MLP}), a novel feature interaction framework for the CTR prediction task. Specially, it  leverages knowledge distillation to consolidate the capacity of learning explicit feature interaction into a main MLP network, while a parallel MLP simultaneously captures implicit feature interactions as a complement. To effectively optimize  the dual-stream MLP architecture, we further design a specific  learning approach with two alignment strategies for enhancing the compatibility of the two MLP components. 
Experiments demonstrate that \textbf{DS-MLP}, though merely a vanilla MLP structure (the final model), can achieve state-of-the-art performance across three widely used  benchmarks, offering a scalable and efficient solution for large-scale recommendation systems. Our code is available at \href{https://github.com/RUCAIBox/DS-MLP}{\textcolor{teal}{https://github.com/RUCAIBox/DS-MLP}}.

\end{abstract}

\setcopyright{cc}
\setcctype{by}
\acmJournal{TKDD}
\acmYear{2026} \acmVolume{1} \acmNumber{1} \acmArticle{}
\acmMonth{1} \acmDOI{10.1145/3819238}

\ccsdesc[500]{Information systems~Recommender systems}

\keywords{CTR Prediction; Knowledge Distillation; Dual-Stream Model;
Recommender Systems }

\maketitle

\section{Introduction}

Click-through rate~(CTR) prediction has long been a critical task in ecommerce platforms such as Amazon, Netflix and  Yelp. 
The key to achieving good performance is to accurately learn the intrinsic relationship among different features,  so that the model can extrapolate well in various feature interaction cases. 
In the literature, a number of approaches have been proposed to capture the complex feature interactions, such as factorization machines~(FM)~\cite{rendle2010factorization}  and deep cross networks~\cite{wang2017deep}.

Typically, mainstream approaches~\cite{guo2017deepfm,cheng2016wide,yu2020deep,song2019autoint} learn the feature interaction relationships in an ensemble way: they use the high-order multiplicative vector operations (\eg dot-product in FMs) to learn the \emph{explicit} feature interactions, and meanwhile feed all features into an MLP component for capturing  the \emph{implicit} interaction characteristics.  
Despite the notable achievements, these approaches are nonetheless plagued by the following two inherent limitations:

$\bullet$ \emph{High complexity}. Most approaches directly enumerate all feature interactions within a pre-defined order.
In real-world scenarios, the interaction signals are normally associated with thousands of fields and millions of instances, which can even reach a much larger scale for large enterprises. 
Due to the exponentially increased feature combinations, modeling high-order interaction relationships in such a large-scale feature space is highly complex.
Moreover, as the model architecture becomes more intricate, it would be more difficult to effectively optimize the model parameters and identify the optimal hyperparameters. 

$\bullet$ \emph{Imbalanced fusion}.   
Prior work~\cite{lian2018xdeepfm,wang2023towards,wang2021dcn,yu2020deep} often combines the outputs of explicit and implicit modules by directly adding them to produce the final result. 
Since the structures of explicit and implicit components are highly different, it may lead to discrepancies in model outputs. A possible side effect is that one module's output becomes the dominating factor in certain situations, overshadowing the output from the other component. 
Actually, the numerical range of the output of explicit feature interactions would be enlarged as the number of involved features increases~\cite{mao2023finalmlp}, while the output of implicit feature interaction tends to diminish. 
In this case, the explicit component would finally dominate the final prediction results bypassing the implicit component. 

Considering the above limitations, several approaches propose adopting a similar architecture to model these two kinds of feature interactions. For example, MLP components are combined to simplify and unify the prediction framework~\cite{mao2023finalmlp}.  
However, it still relies on additional efforts and computational costs in feature selection to maintain   performance. Furthermore, these studies lack effective design in interaction mechanism, so that the MLP component cannot well learn explicit feature interactions, result in limited expressive capacity for the CTR task.

To address these issues, in this paper, we propose a novel feature interaction framework \textbf{D}ual-\textbf{S}tream \textbf{MLP}~(\textbf{DS-MLP}), which is \emph{simple}, \emph{efficient} yet \emph{capable}, for the CTR prediction task. For simplicity, our approach is developed based on generic MLP components as in existing studies~\cite{mao2023finalmlp}. 
Instead of directly combining these MLP components, we gradually extend the network architecture by taking the optimization procedure: \emph{distillation} $\rightarrow$ \emph{alignment} $\rightarrow$ \emph{overall optimization}.  Specifically, we start with an initial MLP component (called \emph{main MLP}) and employ knowledge distillation to train it by learning from more powerful teacher models. In this process, we empirically find that the main MLP mainly learns explicit interaction relationships due to the imbalanced issues in the teacher model. Considering this problem, we further incorporate another auxiliary MLP component (called \emph{parallel MLP}) to enhance the learning of implicit feature interactions. 
To reduce the discrepancies between the two MLP components, we further propose two new alignment strategies that align both the hidden states (with \emph{batch normalization}) and prediction results (with \emph{direct task supervision}).
Such an alignment approach can effectively integrate the two MLP components, to make them focus on learning the explicit and implicit feature interactions accordingly. 
 The contributions of this paper are summarized as follows:

$\bullet$ We propose a new feature interaction framework, \emph{DS-MLP}, for  CTR prediction based on dual-stream MLP architecture. Specifically, only MLP components with very few layers are involved in the final model (\emph{simple} and \emph{efficient}), while it can achieve better or comparable performance as complicated models (\emph{capable}).  

$\bullet$ We design a gradual training procedure that consists of distillation, alignment and overall optimization, which can effectively optimize the two involved MLP components for CTR prediction. In particular, the main and parallel MLPs can well capture explicit and implicit feature interactions, respectively.

$\bullet$
We conduct extensive experiments across three CTR prediction benchmarks.
DS-MLP consistently outperforms all other CTR models, showing the effectiveness of our approach. In addition, it has low latency in dealing with large datasets, making it highly scalable in real-world application scenarios.

\section{Preliminaries} \label{sec:pre}
 
Click-through rate~(CTR)  prediction seeks to estimate the probability that a user will click on an item.
This task is conventionally framed as a binary classification problem, where the model predicts a binary label (\ie \emph{positive} (clicked) or \emph{negative} (non-clicked)).  Formally, an instance is denoted as an input-label pair $\langle \bm x, y \rangle$, {where the input $\bm x = \{ x_1,  x_2,..., x_m \}$}  encapsulates a diverse set of contextual features, comprising user attributes (\eg user ID, gender), item profiles (\eg item ID, category) as well as interaction context (\eg device type, timestamp), while the corresponding label $y\in \{0, 1\}$ indicates actual click behavior. Given a CTR prediction model $f(\cdot)$, it takes $\bm x$ as input and outputs a predicted label $\hat{y}$.

The optimization of model $f(\cdot)$ tpically involves minimizing the binary cross-entropy~(BCE) loss across all training samples, calculated between the true labels and model predictions. This loss function is formally expressed as:  
 \begin{equation}
  \label{equ:ctr}
    \mathcal{L}_{CTR} = -\frac{1}{N} \sum_{i=1}^{N}\bigg(y_i\log(\hat{y}_i)+(1-y_i)\log(1-\hat{y}_i)\bigg). 
  \end{equation}
To improve the prediction accuracy, various CTR prediction methods~\cite{wang2017deep,blondel2016higher,chen2021enhancing} have been proposed in the literature. In general, these methods mainly focus on modeling high-order feature interaction across the context features (encoded in the vector $\bm{x}$), such as xDeepFM~\cite{lian2018xdeepfm} and DCNv2~\cite{wang2021dcn}, thus resulting in increasing computational complexity. In our approach, we take these capable yet complex models as teacher models to guide the learning of lightweight student models via knowledge distillation.

\section{Methodology}

\begin{figure*}[h]
  \centering
  \includegraphics[width=0.98\textwidth]{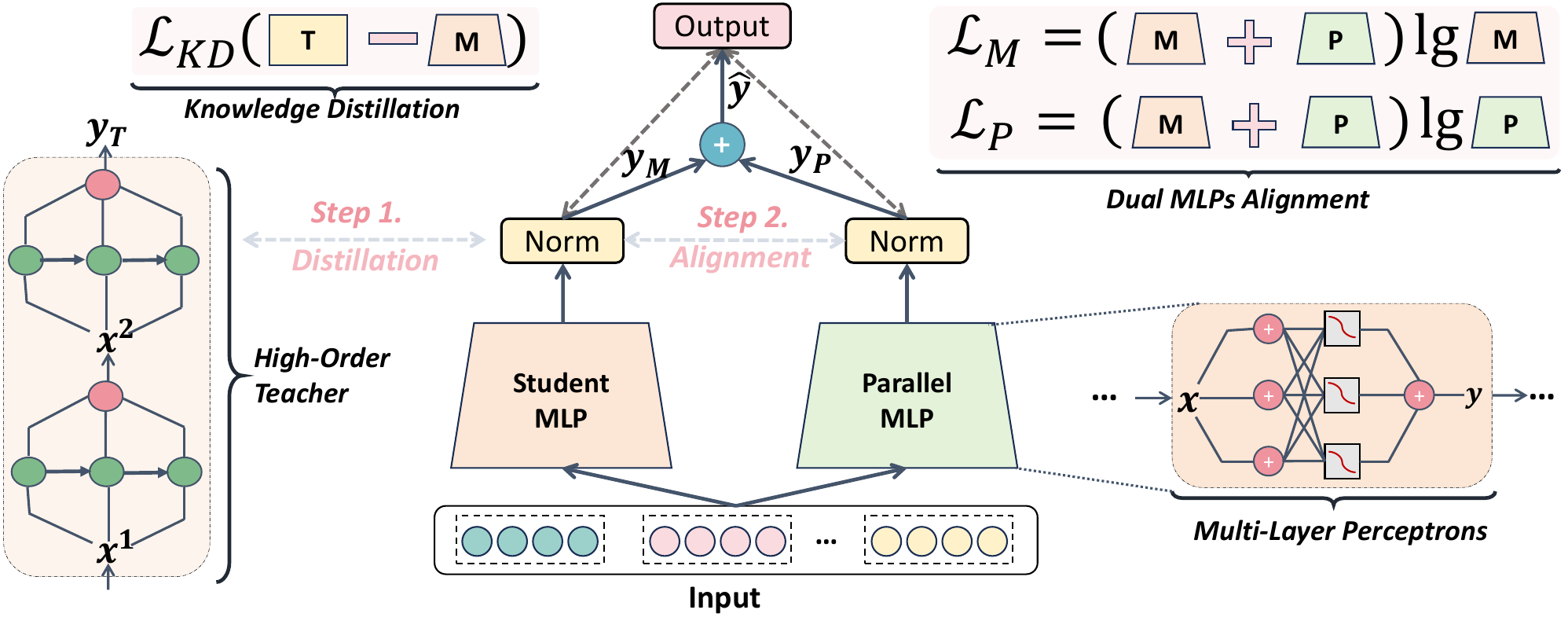}
  \caption{The architecture of our proposed DS-MLP. Overall, it is trained with two main stages, namely knowledge distillation and fine-tuning. }
  \label{fig:framework}
\end{figure*}

In this section, we present the proposed Dual-Stream MLP approach, named  \textbf{DS-MLP} (Figure~\ref{fig:framework}) for the CTR prediction task. Different from existing studies that couple or design intricate network structures, we seek a more simplified,  efficient yet capable CTR model. Instead of simply combining MLP components, we first train a capable \emph{main MLP} component by knowledge distillation (Section~\ref{sec:kd}), and further compensate its capacity on feature interaction learning by incorporating another \emph{parallel MLP} component (Section~\ref{sec:fine-tuning}). 
Next, we introduce the DS-MLP in detail.

\subsection{Universal Feature Interaction Learning via Knowledge Distillation}
\label{sec:kd}
In this section, we explore the role of knowledge distillation in enhancing model performance. We begin by discussing the characteristics of explicit and implicit feature interaction structures and then examine how to design distillation methods to effectively achieve knowledge transfer.

\subsubsection{Capable Feature Interaction Learner as the Teacher (T)}\label{sec:teacher}
Existing state-of-the-art methods predominantly adopt an explicit-implicit integrated way to characterize the complex interaction relations among the context features. 
In this approach, explicit interaction learning is developed based on the concrete feature combinations, while implicit  interaction learning is conducted via deep feature fusion. Since explicit and implicit components are modeled by separate components, such an architecture is often called \emph{dual-stream architecture}. Following this convention, we use the term ``\emph{stream}'' to denote a separate component that contributes to the final output based on given input in a large neural network. 
In our approach, we adopt the capable GDCN~\cite{wang2023towards} model as the teacher model, since it can achieve very impressive results on mainstream CTR benchmarks.  
Specifically, it employs the gated deep cross network~(GCN) as the explicit component and another DNN component as the implicit component, and the final prediction is produced by combining the outputs of the two components, defined as follows:   

\begin{equation}
\phi_{\text{GDCN}}(\bm{x}) = \phi_{\text{GCN}}(\bm{x}) + \phi_{\text{DNN}}(\bm{x}). 
\end{equation}
The core of GDCN lies in the component GCN, which is designed with a more intricate structure for learning explicit feature interactions. 
At each interaction layer, it further utilizes the gated operation to enhance the cross feature interactions~\cite{wang2023towards}, defined as follows:  
\begin{equation}
\label{equ:gcn}
\bm{x}_{l+1}=\underset{\text{Feature}\,\,\text{Crossing}}{\underbrace{\bm{x}_0\odot \left( \mathbf{W}_{l}^{X}  \bm{x}_l+\bm{b}_l \right) }}\odot \underset{\text{Information}\,\,\text{Gate}}{\underbrace{\sigma \left( \mathbf{W}_{l}^{G} \bm{x}_l \right) }}+\bm{x}_l,\\
\end{equation}
where ``$\odot$'' denotes the Hadamard product,  $\mathbf{W}_{l}^{X}$ and $\mathbf{W}_{l}^{G}$ denote the weights at the $l$-th layer for the feature crossing and gating, respectively. In this equation, the feature crossing part extends the interaction order and the information gate employs the previous interaction result (\ie $\bm{x}_l$) to remove irrelevant information from existing feature interactions. 
Furthermore, $\bm{x}_l$ is added to generate the next-order interaction.  In this way, we can increase the number of interaction layers to model higher-order explicit feature interactions, to achieve improved prediction results. 

\subsubsection{Simple MLP as the Student (S)}
In our approach, we consider using the simple MLP architecture to build the student model. The reasons are twofold. First, MLP is a universal unit to build  neural networks of strong expressive capacity, and it theoretically can approximate various complex function relations~\cite{cybenko1989approximation}. Second, existing dual-stream models typically adopt MLP to capture the implicit feature interaction, so that using MLPs as the student model is of potential advantage in model architecture. 
Specifically, MLP is developed based on layerwise information propagation by incorporating linear transformation and nonlinear activation as follows: 
\begin{equation}
\begin{aligned}
    \bm{x}^k &= \sigma (\mathbf{W}^{(k)}\bm{x}^{(k-1)} + \bm{b}^k),
\end{aligned}
\end{equation}
where $k$ represents the layer number, $\sigma$ is an activation function, and $\bm{x}_k$ is the output of the $k$-th layer. The MLP architecture models implicit feature interactions by iteratively stacking more layers,  enabling the model to learn high-order implicit feature interactions. 
In our approach, we aim to keep a  lightweight student model, and set the layer number to three.

\subsubsection{Knowledge Distillation (T $\rightarrow$ S)}
\label{subsec:KD}
After setting the teacher (GDCN) and student (MLP) models, we further employ knowledge distillation for knowledge transfer, to obtain a capable MLP for the CTR prediction task. 

 \paratitle{Overall Framework}. In the literature of machine learning, knowledge distillation~(KD)~\cite{hinton2015distilling} has been proposed to transfer the knowledge from a more powerful model (\ie teacher) to another less capable model (\ie student).   
The basic idea of KD is to conduct supervision signal by reducing the discrepancies between the target knowledge (\ie the information to be distilled) from teacher and student based on the given input, which can be formally described as:
\begin{equation}
    \label{equ:KD}
    \mathcal{L}_{KD} = -\frac{1}{N} \sum_{i=1}^{N} L\bigg(\textsf{OP}\big(\mathcal{M}_T(\bm{x}_i)\big), \textsf{OP}\big(\mathcal{M}_S(\bm{x}_i)\big)\bigg),
\end{equation}
where $\textsf{OP}(\cdot)$ is an operator that obtains the target knowledge (\eg logits or hidden state) that is produced by teacher model $\mathcal{M}_T$ or student model $\mathcal{M}_S$ given input $\bm{x}$, and $L(\cdot)$ is the loss function that measures the difference between two inputs.  

\paratitle{KD Instantiation}. In our approach, we set $\mathcal{M}_T$ to be a more powerful CTR model, \eg GDCN introduced in Section~\ref{sec:teacher}, and $\mathcal{M}_S$ to be a three-layer MLP component. We refer to the MLP as \emph{main MLP}, as it will be enhanced by another auxillary MLP component (Section~\ref{sec:fine-tuning}).  
For knowledge distillation, we consider distilling the logits from the teacher network to guide the learning of the simple MLP network. Furthermore, we adopt the cross entropy loss for $L(\cdot)$ in Eq.~\eqref{equ:KD}. In order to alleviate the logit discrepancy between teacher and student models,  we further incorporate a temperature coefficient to scale the logits. The final optimization objective of the knowledge distillation stage is given as follows:
  \begin{equation}
  \label{equ:KD stage}
    \mathcal{L}=   \mathcal{L}_{CTR}+ \lambda\mathcal{L}_{KD},
  \end{equation}
where  $\lambda$ is a hyperparameter to balance the weights of these two loss functions.
Note that our KD framework can be generally extended to other teacher model designs, while we adopt MLP as the student model due to its generality, lightweight nature, and expressiveness.

\subsection{Capacity Compensation via Dual MLPs}
\label{sec:fine-tuning}
After knowledge distillation, the performance of the student model can be largely improved by learning from the teacher model. Next, we further discuss how to enhance its performance by incorporating a new dual-MLP architecture with a specific training strategy.  

\subsubsection{Dual-stream MLP Architecture}  In our preliminary experiments, we empirically find that, given the dual-stream teacher model GDCN (Section~\ref{sec:teacher}),
the student model mainly focuses on learning the behavior of the explicit component from the teacher model. This is due to an imbalanced fusion of explicit and implicit feature interactions, where the explicit feature crossing dominates the final prediction. As a result, the student model neglects the implicit feature interactions, which might limit its prediction performance.

\paratitle{Component Overwhelming}. By inspecting the intermediate results, a possible reason is that in GDCN, explicit feature interactions are modeled via Hadamard product (Eq.~\eqref{equ:gcn}), so that the numeric range tend to be enlarged for high-order feature interactions. As a comparison,  implicit feature interactions are captured by a standard MLP, whose outputs are typically bounded by activation functions (\eg ReLU) and do not exhibit such exponential growth. As a result, since the teacher’s final prediction is the sum of the explicit and implicit components, the output of the explicit component can sometimes overwhelm the implicit component, causing it to dominate the final prediction.
During knowledge distillation, the student MLP learns to match the teacher's final logits via $\mathcal{L}_{\mathrm{KD}}$ (Eq.~\eqref{equ:KD stage}), primarily focusing on the dominant explicit component. As a result, the student underperforms in modeling implicit interactions.

\paratitle{Enhancement with Parallel MLP}. To address this issue, 
we consider extending the single MLP network to  a dual-stream MLP architecture by incorporating another MLP component. To discriminate the two MLP components, we refer to  the original MLP that serves as the student model as \emph{main MLP}, while the additionally incorporated MLP as \emph{parallel MLP}.   
Similar to dual-stream models~\cite{huang2019fibinet,lian2018xdeepfm}, the parallel MLP component mainly captures the knowledge from the implicit feature interaction, to compensate the performance of the main MLP component. Although dual MLP components have been explored for CTR prediction~\cite{guo2017deepfm,song2019autoint}, they often simply treat both components as equal, while we employ the auxiliary component to enhance the main MLP component. The prediction function of our approach can be formulated as:
\begin{equation}
\label{equ:sum}
\hat{y}_{dual} = \sigma\left(\frac{1}{2}\big(y_M + y_P\big)\right),
  \end{equation}
where $y_M$ and $y_P$ denote the logits of the main and parallel MLP components, respectively. 

\subsubsection{Dual MLPs Alignment}\label{subsec:align} Since these two MLP components in our approach are trained in different ways (only the first MLP component is taught by KD), they tend to produce very divergent outputs, thus hindering the final prediction performance.  Considering this issue, we propose two alignment strategies to combine their outputs in a compatible way, namely hidden state alignment and prediction alignment.

\paratitle{Hidden State Alignment}.  To achieve this goal, we first apply batch normalization~(BN) to the hidden layers of both MLP components, formally given as:
\begin{equation}
\hat{x}_i^m = \frac{x_i^m - \mu_\mathcal{B}^m}{\sqrt{(\sigma_\mathcal{B}^m)^2 + \epsilon}}, \quad i \in \{1, 2, \dots, N\}
\end{equation} 
where $\mu_B^m$ represents the mean of the \(m\)-th feature within the mini-batch, and \(\epsilon\) is a small constant used to prevent division by zero due to an extremely small variance \((\sigma_{\mathcal{B}}^m)^2\).
By regularizing along the batch dimension, the outputs of the dual streams would become more stable and have similar numeric range. Consequently, the logits values obtained through linear mapping are more robust. This can alleviate the issue that the final output is dominated by some module in Eq.~\eqref{equ:sum}.

\paratitle{Prediction Alignment}.  Apart from the differences in learned feature interaction patterns, a simple combination of logits tend to result in a significant  discrepancy in the outputs of both components. The major reason is that the two components are jointly optimized to 
fit the ground-truth label, while they might be unaware of the direct task goal as independent predictors. 
 Therefore, after knowledge distillation, we further introduce the original CTR loss to fine-tune both components as follows: 
\begin{equation}
\begin{aligned}
\label{equ:align}
    \mathcal{L}_{M} &= -\frac{1}{N} \sum_{i=1}^{N}\bigg(y_i\log(\hat{y}_{M,i})+(1-y_i)\log(1-\hat{y}_{M,i})\bigg), \\
    \mathcal{L}_{P} &= -\frac{1}{N} \sum_{i=1}^{N}\bigg(y_i\log(\hat{y}_{P,i})+(1-y_i)\log(1-\hat{y}_{P,i})\bigg),
\end{aligned}
\end{equation}
where $\mathcal{L}_{M}$ and $\mathcal{L}_{P}$ represent the alignment losses for the main and parallel MLP components, respectively. This training loss helps both components adjust their predictions, to be compatible with each other. In addition, it can enhance the task capacity for each component independently, which also alleviates the overwhelming issue of the stronger component.  

\subsubsection{Overall Optimization}. Overall, our approach follows a procedure: \emph{distillation} $\rightarrow$ \emph{alignment} $\rightarrow$ \emph{overall optimization}. In the above content, we have described the distillation method (Section~\ref{subsec:KD}) for learning from strong teacher models and the alignment method (Section~\ref{subsec:align}) for combining both MLP components. Next, we introduce how to conduct joint optimization towards the task goal. 
We define the final optimization objective of the fine-tuning stage for dual-stream MLP architecture:
\begin{equation}
\label{equ:finetuning}
    \mathcal{L}=  \mathcal{L}_{CTR}+ \alpha \mathcal{L}_{M}+ \beta\mathcal{L}_{P},
  \end{equation}
where $\mathcal{L}_{CTR}$ is the main loss (Eq.~\eqref{equ:ctr}) for the CTR task with the prediction from both components as in Eq.~\eqref{equ:sum}, and $\alpha\mathcal{L}_{M}$, and  $\beta\mathcal{L}_{P}$ are the prediction alignment loss functions defined in Eq.~\eqref{equ:align}. 
By adjusting the values of $\alpha$ and $\beta$, we can control the degree of alignment between the dual streams. 
It aims to ensure that their outputs are balanced: neither too extreme nor too identical for preserving the distinct characteristics of each MLP’s learned feature interactions. 
Another note is that the parallel MLP does not reuse the teacher's feature embeddings, since it does not aim to imitate the behavior of the teacher model. We learn the feature embeddings from scratch to enhance its expressive capacity in capturing implicit feature interactions. 

\paragraph{Effectiveness.} Our method effectively leverages the complementary nature of explicit and implicit feature interactions. Through knowledge distillation, the student MLP captures the explicit feature crossing learned by the teacher network, while the parallel MLP learns the complementary implicit feature interactions. Unlike traditional dual-stream models that simply sum the explicit and implicit components, which often results in the explicit component dominating and undermining the implicit interactions, our method ensures a more balanced integration. By aligning the outputs of both MLPs via the alignment loss (Eq.~\eqref{equ:align}), we prevent one component from overwhelming the other. This alignment allows the model to fully utilize both types of feature interactions, thereby improving overall predictive capacity without sacrificing either explicit or implicit interactions.

\subsection{Algorithm Overview}

Algorithm~\ref{alg:dsmlp} presents the overall training procedure of our proposed DS-MLP framework for CTR prediction, which consists of two complementary stages: knowledge distillation and dual-MLP fine-tuning.

\paragraph{Stage I: Universal Feature Interaction Learning via Knowledge Distillation.}
In the first stage, the main MLP branch is trained under the supervision of a powerful teacher model. For each minibatch, the frozen teacher network produces predictions that serve as soft targets. Simultaneously, the main and parallel MLP branches generate their respective outputs, which are aggregated to obtain the overall prediction. Knowledge distillation is performed on the hidden representations extracted from the branches, encouraging the student network to mimic the teacher’s behavior. In parallel, the standard CTR prediction loss is computed based on the aggregated output. The main MLP is then updated by jointly minimizing the distillation loss and the CTR loss, enabling it to effectively capture explicit feature interactions in a compact manner.

\paragraph{Stage II: Capacity Compensation via Dual MLPs (Finetuning).}
After the initial distillation stage, we finetune the dual-MLP architecture to further enhance the model’s capacity. Hidden states from the dual streams are first normalized with batch normalization to improve stability and mitigate discrepancies. We then compute branch-wise CTR losses to construct a prediction-alignment loss that enforces consistency between the two branches. The overall loss combines the CTR loss of the aggregated prediction with the alignment loss, guiding the network to achieve a balanced modeling of explicit and implicit interactions. Through this finetuning process, the parallel MLP complements the main branch, thereby improving both robustness and generalization.

\paragraph{Final Prediction.}

After training, the DS-MLP produces the final prediction by aggregating the outputs of both the main and parallel MLP branches. 
This approach effectively combines the main branch's distilled knowledge of explicit feature interactions with the parallel branch's capacity for capturing implicit ones. Our two-stage training, unified by an alignment objective, ensures these complementary strengths are balanced, yielding a final model that is both lightweight and highly expressive.

\begin{algorithm}[t]
    \caption{Training Procedure of DS-MLP}
    \label{alg:dsmlp}
    \begin{algorithmic}[1]
        \State \textbf{Input:} Training data $\mathcal{D}=\{(x_i, y_i)\}_{i=1}^N$; teacher model $\mathcal{T}$; main MLP $\mathcal{M}$; parallel MLP $\mathcal{P}$
        \State \textbf{Output:} Final prediction $y_{pred}$
        \State \textbf{Part I: Universal Feature Interaction Learning (Knowledge Distillation)}
        \For{each minibatch $(x, y) \in \mathcal{D}$}

            \State Generate teacher output: $p_T = \mathcal{T}(x) \quad \textbf{(frozen)}$
            \State Produce branch outputs: $p_M = \mathcal{M}(x)$, $p_P = \mathcal{P}(x)$
            \State Aggregate branchs prediction: $p = \gamma p_M + (1-\gamma) p_P$ \quad (with $\gamma=0.5$ in practice)
            \State Extract hidden states $(\mathcal{M}_T,\mathcal{M}_S)$ and compute distillation loss:\par
            \hspace{1.2em}$\mathcal{L}_{KD} = \text{KL}(\mathcal{M}_T \parallel \mathcal{M}_S)$ \hfill $\blacktriangleright$ Eq.~\eqref{equ:KD}
            \State Compute CTR loss: $\mathcal{L}_{CTR} = \text{BCE}(p, y)$ \hfill $\blacktriangleright$ Eq.~\eqref{equ:ctr}
            \State Update $\mathcal{M}$ via joint optimization:\par
            \hspace{1.2em}$\mathcal{L} = \mathcal{L}_{CTR} + \lambda \mathcal{L}_{KD}$ \hfill $\blacktriangleright$ Eq.~\eqref{equ:KD stage}
        \EndFor

        \State \textbf{Part II: Capacity Compensation via Dual MLPs (Finetuning)}
        \For{each minibatch $(x, y) \in \mathcal{D}$}

            \State Generate branch hidden output: $h_M = \mathcal{M}(x)$, $h_P = \mathcal{P}(x)$
    \State Hidden State Alignment: 
    $\hat{h}_M = BN(h_M)$, $\hat{h}_P = BN(h_P)$
            \State Aggregate branchs prediction:
            $p = \gamma p_M + (1-\gamma) p_P$ \quad (with $\gamma=0.5$ in practice)
            \State Compute branch-wise task losses:  \par
            \hspace{1.2em}$\mathcal{L}_{M} = \text{BCE}(p_M, y)$, \quad $\mathcal{L}_{P} = \text{BCE}(p_P, y)$ \hfill $\blacktriangleright$ Eq.~\eqref{equ:align}
            \State Define prediction alignment loss: $\mathcal{L}_{align} = \alpha \mathcal{L}_{M}+ \beta\mathcal{L}_{P},$
            \State Optimize with joint objective:\par
            \hspace{1.2em}$\mathcal{L} = \mathcal{L}_{CTR} + \mathcal{L}_{align}$ \hfill $\blacktriangleright$ Eq.~\eqref{equ:finetuning}
        \EndFor
        \State \textbf{Final Prediction:} $y_{pred} = p$
    \end{algorithmic}
\end{algorithm}

\subsection{Discussion}
\label{method:dis}
In this part, we summarize the key features of our method DS-MLP.

$\bullet$~\emph{Compatibility}. A defining feature of our approach is its teacher-agnostic design. DS-MLP can flexibly incorporate any state-of-the-art CTR model as a teacher, thereby ensuring that the framework remains future-proof and continuously adaptable. As more powerful models emerge, they can be seamlessly adopted as new teachers to transfer their enhanced capabilities into the efficient student architecture.
 In addition, the exclusive reliance on an MLP backbone renders DS-MLP inherently deployment-friendly. The simple and homogeneous structure of MLPs is highly optimized in modern deep-learning libraries and benefits from extensive hardware acceleration, which facilitates straightforward implementation and maintenance in low-latency industrial environments. 

$\bullet$~\emph{Scalability}. The framework exhibits strong scalability as a direct consequence of its MLP-based foundation. MLPs naturally scale to massive datasets owing to their inherent parallelizability and structural simplicity. When larger or more complex data must be handled, the capacity of the student model can be increased simply by widening the hidden layers or deepening the network. This property provides practitioners with a straightforward and predictable means of improving model performance, while avoiding the extensive architectural redesigns typically required for scaling more specialized or complex models.

$\bullet$~\emph{Complexity}. 
Our approach employs several effective strategies to ensure the model efficiency.
First, unlike previous dual-stream models, we use MLP to model both explicit and implicit feature interactions. It is more efficient for modeling high-order interactions, and can also avoid the imbalanced fusion issue caused by the exponentially increased explicit interactions~\cite{wang2023towards,zhu2023final}.  
Secondly, our architecture is decoupled, which means the student MLP can be integrated with multiple parallel MLPs. As such, it greatly reduces the difficulty of parameter tuning and enhances the flexibility for adapting to various industrial scenarios. 

The overall comparison of our approach with existing methods is presented in Table~\ref{tab:discussion}.

\begin{table}[!h]
\centering
    \caption{Comparison of different two-stream methods. ``SE'' denotes that the model only uses single embedding table.}
    \label{tab:discussion}
    \begin{tabular}{l|cccc|c}
    \toprule
    Methods	& {Explicit} &	{Implicit} & {Alignment} & SE & Structure\\
		\hline
  GDCN &	\textcolor{teal}{\CheckmarkBold} &	\textcolor{teal}{\CheckmarkBold} &	 \textcolor{purple}{\XSolidBrush} & \textcolor{teal}{\CheckmarkBold} & Cross Network + MLP\\
FinalMLP	&	 \textcolor{purple}{\XSolidBrush} &	\textcolor{teal}{\CheckmarkBold} &	  \textcolor{purple}{\XSolidBrush}  &  \textcolor{purple}{\XSolidBrush} & Selection Gate + MLP\\
Final	&	\textcolor{teal}{\CheckmarkBold} &	\textcolor{purple}{\XSolidBrush} &	\textcolor{teal}{\CheckmarkBold} & \textcolor{teal}{\CheckmarkBold} & Deep Cross Network\\
DS-MLP	&	\textcolor{teal}{\CheckmarkBold} &	\textcolor{teal}{\CheckmarkBold} &	\textcolor{teal}{\CheckmarkBold} & \textcolor{teal}{\CheckmarkBold} & MLP Only \\
\bottomrule
\end{tabular}

\end{table}

\subsection{Theoretical Analysis}

In this section, we theoretically justify that a ReLU-based MLP can recover the explicit feature interactions encoded in a cross network (\eg GDCN) via knowledge distillation. Using the notations from Sec.~\ref{sec:kd}, the proof proceeds by demonstrating that the student $\mathcal{M}_S$ matches the teacher $\mathcal{M}_T$ mapping over inputs $\bm{x}\in\mathcal{X}\subset\mathbb{R}^d$, both in output approximation and underlying interaction mechanics.

\paratitle{Approximation via Knowledge Distillation.} The GDCN teacher $\mathcal{M}_T: \mathcal{X} \to [0,1]$ is a continuous mapping on a compact domain $\mathcal{X}$. The student $\mathcal{M}_S$ is a ReLU-based MLP. By the Universal Approximation Theorem \citep{hornik1989multilayer}, the hypothesis space $\mathcal{F}_S$ of the student is dense in the space of continuous functions. Hence, for any $\varepsilon>0$, there exists $\mathcal{M}_S^*\in\mathcal{F}_S$ that approximates $\mathcal{M}_T$ arbitrarily well in the $L^2(\mathcal{D})$ norm. During distillation, minimizing the expected squared error $\mathcal{L}_{\mathrm{KD}} = \mathbb{E}_{\bm{x}\sim\mathcal{D}}[(\mathcal{M}_T(\bm{x})-\mathcal{M}_S(\bm{x}))^2]$ performs empirical $L^2$ regression toward $\mathcal{M}_T$. Assuming standard overparameterized optimization \citep{du2019gradient, zou2018stochastic}, the learned student $\hat{\mathcal{M}}_S$ converges with bounded excess risk, guaranteeing it tightly approximates the teacher:
\begin{equation}
\|\hat{\mathcal{M}}_S-\mathcal{M}_T\|_{L^2(\mathcal{D})}^2 \lesssim \inf_{\mathcal{M}_S\in\mathcal{F}_S}\|\mathcal{M}_S-\mathcal{M}_T\|_{L^2(\mathcal{D})}^2 + \mathcal{O}\!\left(\frac{1}{\sqrt{n}}\right). 
\end{equation}

\paratitle{Feature Interaction Recovery and Generalization.} To verify that the student internalizes the teacher's interaction sensitivity rather than merely memorizing outputs, we analyze its second-order behavior. The interaction strength between features $i$ and $j$ is measured by the expected absolute mixed partial derivative:
\begin{equation}
I_{ij}(f) = \mathbb{E}_{\bm{x}\sim\mathcal{D}}\left[ \left| \frac{\partial^2 f}{\partial x_i \partial x_j}(\bm{x}) \right| \right].
\end{equation}
Assume that both $\mathcal{M}_T$ and the student network $\hat{\mathcal{M}}_S$ are sufficiently smooth (\eg can be approximated by $C^2$ functions). Under such regularity, standard results in approximation theory \citep{pinkus1999approximation} guarantee the existence of networks that simultaneously approximate a function and its derivatives. However, to link the $L^2$ function approximation obtained from distillation to derivative approximation, we note that if $\|\hat{\mathcal{M}}_S-\mathcal{M}_T\|_{L^2(\mathcal{D})} \leq \varepsilon$, then by interpolation inequalities in Sobolev spaces (or by the fact that neural networks with bounded parameters have bounded Lipschitz constants), we can further deduce that the Hessians satisfy $\|\nabla^2\hat{\mathcal{M}}_S-\nabla^2\mathcal{M}_T\|_{L^2(\mathcal{D})} = \mathcal{O}(\varepsilon^\alpha)$ for some $\alpha>0$, provided the network architecture and training impose sufficient smoothness (\eg via weight decay or implicit regularization). Consequently, the interaction difference is bounded by:
\begin{equation}
\begin{aligned}
|I_{ij}(\hat{\mathcal{M}}_S) - I_{ij}(\mathcal{M}_T)|
&\lesssim
\mathbb{E}_{\bm{x}}\!\left[
\|\nabla^2\hat{\mathcal{M}}_S(\bm{x})
- \nabla^2\mathcal{M}_T(\bm{x})\|_F
\right] \\
&\lesssim
\|\nabla^2\hat{\mathcal{M}}_S
- \nabla^2\mathcal{M}_T\|_{L^2(\mathcal{D})}
\to 0 \quad \text{as } \varepsilon \to 0.
\end{aligned}
\end{equation}
This alignment of second-order derivatives confirms that the student inherits the explicit high-order feature interactions of the cross network under mild regularity conditions.

Furthermore, the $L^2$ approximation guarantees downstream task performance. For a loss function $\ell$ that is $L$-Lipschitz in its first argument, the difference in expected risk $R(f)=\mathbb{E}_{\bm{x},y}[\ell(f(\bm{x}),y)]$ is bounded by:
\begin{equation}
|R(\hat{\mathcal{M}}_S) - R(\mathcal{M}_T)| \leq L\,\mathbb{E}_{\mathcal{D}}[|\hat{\mathcal{M}}_S-\mathcal{M}_T|] \leq L\,\|\hat{\mathcal{M}}_S-\mathcal{M}_T\|_{L^2(\mathcal{D})}.
\end{equation}
In summary, knowledge distillation drives the MLP to approximate the teacher not only in its output but also at the Hessian level, thereby reconstructing explicit feature interactions. Thus, replacing complex interaction modules with a distilled plain MLP is theoretically justified under appropriate smoothness and optimization assumptions.

\section{Experiments}

In this section, we demonstrate the effectiveness and transferability of the proposed DS-MLP approach.

\subsection{Experimental Setup}
We introduce the experimental settings, including the datasets,
baseline approaches, metrics, and hyper-parameter details.

\subsubsection{Datasets}
We conduct experiments on three public datasets for recommender evaluation, including: Criteo\footnote{https://www.kaggle.com/c/criteo-display-ad-challenge/data}, Avazu\footnote{https://www.kaggle.com/c/avazu-ctr-prediction/data}, Movielens\footnote{https://grouplens.org/datasets/movielens}. 
The splitting and pre-processing proceducheng2020adaptiveres follow the AFN~\cite{cheng2020adaptive} work.  
The statistics of datasets are summarized in Table~\ref{tab:datasets}.

\begin{table}[h]
  \caption{The statistics of datasets.} 
  \label{tab:datasets}
  \begin{tabular}{c|ccc}
    \toprule
    \textbf{Dataset}& $\#$ Instances & $\#$ Fields & $\#$ Features \\
    \hline \hline
    Criteo & 45,840,617 & 39 & 2,086,936 \\
    Avazu & 40,428,967 & 22 & 1,544,250 \\
    MovieLens& 2,006,859 & 3 & 90,445\\
  \bottomrule
\end{tabular}

\end{table}

$\bullet$ Criteo is a widely-used benchmark dataset for CTR prediction, which contains about one week of click-through data for display advertising. It has 13 numerical feature fields and 26 categorical feature fields.

$\bullet$ Avazu originates from the 2014 Avazu Click-Through Rate Prediction contest and contains about 10 days of mobile advertising records. Each instance indicates whether a user clicked the displayed ad, and the dataset includes 23 categorical features.

$\bullet$ Movielens contains users' tagging behaviors on movies. For personalized tag recommendation, each tagging record, consisting of a user ID, a movie ID, and a tag, is converted into a feature vector as input. The target indicates whether a specific tag was assigned by the user to the corresponding movie.

\subsubsection{Compared Methods}

We compare the proposed approach with the following baseline methods:

$\bullet$ \textbf{FmFM}~\cite{pan2018field}: it builds upon FwFM by incorporating a field matrix and employing kernel products to model feature interactions. This allows FmFM to capture more flexible and fine-grained relationships between features at the field level.\par
$\bullet$ \textbf{AFM}~\cite{xiao2017attentional}: it improves upon FM by applying an attention mechanism to discriminate the importance of different feature interactions. AFM dynamically assigns weights to each interaction, enabling the model to focus on more informative feature pairs.\par
$\bullet$ \textbf{AFN}~\cite{cheng2020adaptive}: it automatically learns the optimal order of feature interactions from data, overcoming the limitations of fixed-order models. Its core innovation lies in a Logarithmic Transformation Layer that adaptively estimates an exponent for each feature, enabling the model to capture arbitrary-order feature crossings more efficiently. AFN+ further integrates a deep neural network to improve model stability and enhance predictive performance.\par
$\bullet$ \textbf{FiBiNet}~\cite{huang2019fibinet}: it utilizes a Squeeze-and-Excitation (SENet) module together with bilinear feature interactions to learn both the importance of individual features and their pairwise interactions. This dual mechanism allows FiBiNet to adaptively reweight features and interactions based on their contribution to the prediction task.\par
$\bullet$ \textbf{FiGNN}~\cite{li2019fi}: it  models feature interactions by treating features as nodes in a graph. A Graph Neural Network (GNN) is then used to iteratively aggregate information from neighboring nodes, enabling the capture of high-order, structured relationships.\par
$\bullet$ \textbf{PNN}~\cite{qu2017product}: 
it explicitly captures second-order interaction patterns by inserting a dedicated product layer between the embedding and deep network components. This layer computes both inner and outer products of feature embeddings, thereby enriching the signals passed to the subsequent MLP and improving the model’s ability to learn cross-feature patterns.\par
$\bullet$ \textbf{DeepFM}~\cite{guo2017deepfm}: it combines the strengths of FM for modeling low-order explicit feature interactions with a deep neural network (MLP) for capturing high-order implicit interactions, providing a unified framework that leverages both explicit and implicit information.\par
$\bullet$ \textbf{xDeepFM}~\cite{lian2018xdeepfm}: it employs a Compressed Interaction Network (CIN) to generate explicit high-order feature interactions at the vector level. CIN efficiently captures interactions across multiple layers without the need for parameter-heavy fully connected layers.\par
$\bullet$ \textbf{DCNv2}~\cite{wang2021dcn}: it extends the Deep \& Cross Network framework by stacking multiple deep cross layers to model high-order feature interactions, while integrating an MLP to capture implicit interactions and integrates an MLP for implicit interactions.\par
$\bullet$ \textbf{AutoInt}~\cite{song2019autoint}: 
it leverages a multi-head self-attention mechanism to explicitly model feature interactions and assign adaptive importance weights to each pairwise interaction. Its enhanced version, AutoInt+, incorporates a feed-forward network to improve expressiveness and capture more complex non-linear dependencies.\par
$\bullet$ \textbf{GDCN}~\cite{wang2023towards}: it utilizes a Gated Deep Cross Network where a gating mechanism is applied at each cross layer. This allows the network to selectively filter and emphasize important feature interactions at each order, which helps reduce noise from less useful crosses and improves model robustness.\par
$\bullet$ \textbf{FCN}~\cite{li2024FCN}: 
it employs a Fusing Cross Network that integrates both linear and exponential sub-networks. This is further enhanced with a Self-Mask mechanism, which dynamically suppresses redundant cross terms, and a tailored Tri-BCE loss, which improves optimization stability.
This design allows efficient learning of high-order interactions while controlling overfitting and redundancy.\par
$\bullet$ \textbf{FinalMLP}~\cite{mao2023finalmlp}: it introduces stream-specific feature gating and multi-head bilinear fusion modules to enhance the representation power of MLP-based models. This allows selective fusion of different feature streams, improving the capacity to model diverse interaction patterns.\par
$\bullet$ \textbf{Final}~\cite{zhu2023final}: 
it introduces a Factorized Interaction Layer that extends the widely used linear layer to explicitly learn second-order feature interactions. Similar to MLPs, multiple FINAL layers can be stacked into a FINAL block, enabling the model to capture exponentially growing degrees of feature interactions. This design offers a lightweight yet powerful way to model complex combinatorial feature patterns without resorting to excessively deep networks.\par
$\bullet$ \textbf{WuKong}~\cite{zhang2024wukong}:
it introduces a scaling-oriented recommendation architecture built upon stacked factorization machines, coupled with a synergistic upscaling strategy. By progressively increasing depth and width, Wukong captures diverse high-order feature interactions while establishing a scaling law in recommendation modeling.\par
$\bullet$ \textbf{SFG}~\cite{yin2025feature}: it shifts CTR modeling from discriminative feature interaction to supervised feature generation via an encoder–decoder framework. By reconstructing feature embeddings under click supervision, SFG alleviates embedding collapse and information redundancy.\par
$\bullet$ \textbf{ECKD}~\cite{zhu2020ensembled}:
it aggregates knowledge from multiple teacher models into a single student network via knowledge distillation. By combining the strengths of various teacher architectures, ECKD maintains a compact yet powerful student model for efficient inference.\par

The methods mentioned above cover various approaches in the field of CTR prediction. 
Among them, FmFM, and AFM model second-order feature interactions. 
DeepFM, xDeepFM, DCNv2, GDCN, FiBiNet, and AutoInt jointly model explicit and implicit high-order feature interactions, while FCN does so by linearly and exponentially increasing the interaction order. 
Other architectures like Final and FinalMLP are characterized by their dual-stream designs, and ECKD represents a KD-based approach. More recently, SFG models cross-feature dependencies implicitly through a supervised feature generation paradigm, while Wukong captures arbitrary-order interactions via stacked factorization machines with scalable depth and width expansion.
In contrast to the above methods, our DS-MLP approach utilizes knowledge distillation to unify different types of feature interactions and ensures equal utilization of the dual-stream MLP outputs. 

\subsubsection{Evaluation Metrics}

Following previous works~\cite{huang2019fibinet,song2019autoint,zhou2019deep}, we adopt two common ranking metrics in CTR models: AUC and LogLoss. \par
$ \bullet$ \textbf{AUC~\cite{lobo2008auc}}, which represents the probability that a randomly chosen positive example is ranked higher than a randomly chosen negative example.\par
$\bullet$ \textbf{LogLoss~\cite{buja2005loss}}, which measures the accuracy of the predicted probabilities against the actual binary outcomes, with lower values indicating better model performance. 

The \textbf{RelaImpr} metric, as proposed in~\cite{zhou2019deep}, measures the relative improvement with respect to the base model and is defined as follows:
\begin{equation}
    \textit{RelaImpr} = \left(\frac{\textit{AUC}(\textit{measure model}) - 0.5}{\textit{AUC}(\textit{base model}) - 0.5} - 1\right) \times 100\%
\end{equation}

\begin{table}[t]
  \caption{Hyperparameters for the datasets.} 
  \label{tab:hyper}
  \begin{tabular}{c|ccccc}
    \toprule
    \textbf{Dataset}& Teacher Model & Main MLP Size  & $\lambda$  & $\alpha$  & $\beta$  \\
    \hline \hline
    Criteo & GDCN  &600 & 1.0 & 0.8 & 1.2 \\
    Avazu & DCNv2 &900 & 1.0 & 1.0 & 0.6\\
    MovieLens& FinalMLP &1000 & 0.5 & 1.5 & 0.8\\
  \bottomrule
\end{tabular}

\end{table}

\subsubsection{Implementation Details}
We implement our models based on FuxiCTR, a configurable, tunable, and reproducible library for CTR prediction\footnote{https://github.com/reczoo/FuxiCTR}.
To ensure a fair and consistent comparison, we strictly align the parameter settings of all baseline methods with the BARS benchmark\footnote{https://github.com/reczoo/BARS}. In particular, we fix the embedding dimension to 10, adopt a batch size of 4096, and configure the default MLP architecture as [400, 400, 400]. All models are optimized using the Adam optimizer with its default parameters. For ECKD, we reproduce the best-performing configuration reported in the original paper. Specifically, AutoInt+, DCNv2, and xDeepFM are selected as teacher models, and the “soft label + pre-train” strategy is adopted to transfer knowledge into the student. For DS-MLP, we design the student MLP with a larger hidden size of [600, 600, 600] to ensure sufficient capacity for absorbing knowledge from complex teacher architectures, while keeping the parallel MLP size identical to the teacher MLP module. To examine the trade-off between model expressiveness and efficiency, we progressively scale the hidden size of each layer from 1200 down to 100, thereby obtaining a comprehensive view of how parameterization affects performance.

For the distillation loss weight $\lambda$, we explore values in the range [0.1, 0.5, 1, 1.5, 10], while the fine-tuning loss weights $\alpha$ and $\beta$ follow a similar trend, gradually varying from 0.5 to 1.5 with a step size of 0.1. 
The final, best-performing configurations are presented in Table~\ref{tab:hyper}, revealing that the optimal student MLP size is highly dependent on the dataset and teacher architecture. A larger student capacity (\eg width 1000) is necessary to distill knowledge from complex teachers with implicit interactions like FinalMLP, while a more compact student is sufficient for teachers with structured, explicit patterns like DCNv2. Moreover, on large-scale datasets like Criteo, a more modest MLP size (600) proved optimal, striking a crucial balance between learning from the teacher's guidance and avoiding over-reliance on its potential biases.

\begin{table}[htbp]
\renewcommand{\arraystretch}{0.75}
\centering
\caption{Performance comparisons grouped by dataset. Note that a higher AUC or lower Logloss at 0.001-level is significant for CTR prediction.}
\label{tab:overall}
\begin{tabular*}{0.9\columnwidth}{@{\extracolsep{\fill}} ll cccc} 
\toprule
\textbf{Dataset} & \textbf{Model} & \textbf{AUC} & \textbf{Log Loss} & \textbf{RelaImpr} & \textbf{Latency~(ms)} \\
\midrule
\multirow{17}{*}{Criteo} 
 & FmFM & 0.8054 & 0.4464 & - & 435.75 \\
 & AFM & 0.8034 & 0.4477 & -0.65\% & 375.07 \\
 & AFN+ & 0.8141 & 0.4377 & 2.85\% & 38.55 \\
 & FiBiNet & 0.8128 & 0.4390 & 2.42\% & 1018.28 \\
 & FiGNN & 0.8134 & 0.4384 & 2.62\% & 390.47 \\
 & PNN & 0.8133 & 0.4386 & 2.59\% & 84.20 \\
 & DeepFM & 0.8134 & 0.4380 & 2.62\% & 24.58 \\
 & xDeepFM & 0.8136 & 0.4382 & 2.69\% & 182.80 \\
 & DCNV2 & 0.8132 & 0.4389 & 2.55\% & 41.38 \\
 & AutoInt+ & 0.8138 & 0.4380 & 2.75\% & 1262.53 \\
 & GDCN & 0.8137 & 0.4386 & 2.72\% & 45.02 \\
 & FCN & \underline{0.8149} & \underline{0.4368} & 3.11\% & 39.26 \\
 & FinalMLP & 0.8147 & 0.4374 & 3.05\% & 56.71 \\
 & Final & 0.8148 & 0.4369 & 3.08\% & 25.56 \\
  & WuKong & 0.8140 & 0.4388 &  2.82\% & 34.35 \\
   & SFG & 0.8144 & 0.4378 &  2.95\% &  32.78 \\
 & ECKD & 0.8139 & 0.4401 & 2.78\% & 43.15 \\
\cmidrule(lr){2-6} 
 & \textbf{DS-MLP} & \textbf{0.8152} & \textbf{0.4366} & \textbf{3.21\%} & 59.43 \\
\midrule
\multirow{17}{*}{Avazu} 
 & FmFM & 0.7599 & 0.3688 & - & 22.07 \\
 & AFM & 0.7557 & 0.3714 & -1.62\% & 33.75 \\
 & AFN+ & 0.7643 & 0.3673 & 1.69\% & 21.21 \\
 & FiBiNet & 0.7622 & 0.3680 & 0.88\% & 19.60 \\
 & FiGNN & 0.7608 & 0.3686 & 0.35\% & 25.64 \\
 & PNN & 0.7617 & 0.3679 & 0.69\% & 26.30 \\
 & DeepFM & 0.7638 & 0.3675 & 1.50\% & 20.23 \\
 & xDeepFM & 0.7639 & 0.3677 & 1.54\% & 26.31 \\
 & DCNV2 & 0.7627 & 0.3677 & 1.08\% & 18.51 \\
 & AutoInt+ & 0.7634 & 0.3675 & 1.35\% & 36.49 \\
 & GDCN & 0.7648 & 0.3673 & 1.89\% & 21.52 \\
 & FCN & 0.7651 & 0.3666 & 2.00\% & 28.73 \\
 & FinalMLP & 0.7659 & 0.3662 & 2.31\% & 21.03 \\
 & Final & \underline{0.7664} & \underline{0.3659} & 2.50\% & 19.26 \\
  & WuKong & 0.7641 & 0.3683 & 1.62\% & 27.49 \\
   & SFG & 0.7650 & 0.3679 & 1.96\% & 20.19 \\
 & ECKD & 0.7646 & 0.3815 & 1.81\% & 21.13 \\
\cmidrule(lr){2-6}
 & \textbf{DS-MLP} & \textbf{0.7670} & \textbf{0.3657} & \textbf{2.73\%} & 25.89 \\
\midrule
\multirow{17}{*}{MovieLens} 
 & FmFM & 0.9482 & 0.2467 & - & 3.30 \\
 & AFM & 0.9439 & 0.2733 & -0.96\% & 4.62 \\
 & AFN+ & 0.9641 & 0.3010 & 3.55\% & 3.16 \\
 & FiBiNet & 0.9527 & 0.2556 & 1.00\% & 6.62 \\
 & FiGNN & 0.9516 & 0.2573 & 0.76\% & 5.25 \\
 & PNN & 0.9684 & 0.2099 & 4.51\% & 4.93 \\
 & DeepFM & 0.9689 & 0.2116 & 4.62\% & 5.42 \\
 & xDeepFM & 0.9685 & 0.2151 & 4.53\% & 3.78 \\
 & DCNV2 & 0.9690 & 0.2139 & 4.64\% & 3.03 \\
 & AutoInt+ & 0.9688 & 0.2202 & 4.60\% & 3.64 \\
 & GDCN & 0.9685 & 0.2164 & 4.53\% & 3.75 \\
 & FCN & 0.9671 & 0.2076 & 4.22\% & 3.54 \\
 & FinalMLP & 0.9716 & 0.2119 & 5.22\% & 2.31 \\
 & Final & \underline{0.9717} & \underline{0.1978} & 5.24\% & 2.29 \\
  & WuKong & 0.9692 & 0.2136 & 4.68\% & 4.41 \\
   & SFG & 0.9698 & 0.2118 & 4.82\% &  3.02 \\
 & ECKD & 0.9686 & 0.2182 & 4.55\% & 3.31 \\
\cmidrule(lr){2-6}
 & \textbf{DS-MLP} & \textbf{0.9752} & \textbf{0.1971} & \textbf{6.02\%} & 2.65 \\
\bottomrule
\end{tabular*}
\end{table}

\subsection{Overall Performance}
We evaluate DS-MLP against a comprehensive set of baseline methods on three widely used CTR prediction benchmarks. The results, summarized in Table~\ref{tab:overall}, provide a clear and systematic comparison across models of different architectures and design principles. Several key observations can be drawn from these experiments.


$\bullet$~\emph{Effectiveness of High-Order Modeling.} High-order feature interaction models (\eg PNN, DeepFM, and Final) consistently outperform second-order approaches (\eg FmFM and FwFM), confirming the importance of modeling complex feature interactions for CTR prediction. Among these, explicit high-order methods such as xDeepFM, DCNv2, and WuKong outperform implicit methods like FiBiNet, reflecting the advantage of directly modeling interaction structures rather than relying solely on implicit representations. FCN further extends DCNv2 by explicitly capturing both linear and exponential interactions, thereby reducing the dependence on implicit DNN layers. GDCN strengthens this advantage even more by employing gated mechanisms at each interaction order to dynamically filter and emphasize the most informative feature combinations. Meanwhile, SFG models high-order dependencies through supervised feature generation and reconstruction, offering an alternative generative perspective for capturing complex feature interactions. 

$\bullet$~\emph{Strength of MLP-Based Architectures.} 
MLPs are well recognized as universal function approximators. When properly configured, they can substantially improve the performance of both standalone and hybrid architectures. For example, in AutoInt+ and AFN+, an MLP is employed to model implicit feature interactions, while FinalMLP adopts an MLP entirely as its backbone, where feature gating further refines the information flow. These observations demonstrate that, despite their apparent simplicity, MLPs remain a highly competitive and versatile backbone for CTR modeling.

$\bullet$~\emph{Effectiveness of Knowledge Distillation.} The strong performance of ECKD validates the knowledge distillation paradigm. By ensembling the distilled knowledge from a diverse set of teacher networks, it synthesizes their combined strengths, allowing the student model to ultimately surpass the performance of any single teacher.

$\bullet$ Ultimately, our proposed DS-MLP effectively acquires explicit high-order feature interactions from the teacher network through the distillation process. By aligning the main and parallel MLPs, it achieves a coherent integration of both explicit and implicit interaction modeling, thereby enhancing its representational capacity. As a result, DS-MLP consistently surpasses the baseline models and achieves state-of-the-art (SOTA) performance across all three benchmark datasets.

In addition to accuracy, efficiency is a critical factor for real-world deployment. We therefore measure the inference latency of all CTR models under an identical server environment. As reported in Table~\ref{tab:overall}, explicit CTR models (\ie FwFM, FmFM, and AFM) have nearly zero non-embedding parameters, their inference times are considerably long. Parallel architectures such as DeepFM and GDCN demonstrate superior latency–performance trade-offs, achieving improved accuracy at comparatively low latency. In contrast, DS-MLP attains the highest predictive performance while maintaining an inference time comparable to efficient baseline models such as FinalMLP, Final, and DCNv2. This balance between accuracy and efficiency underscores the practicality of DS-MLP for large-scale CTR prediction scenarios.

\begin{table}[t]
  \caption{Compatibility Analysis of DS-MLP.}\label{Compatibility Analysis}
  \begin{tabular}{c|cc|cc|cc}
    \toprule
    \multirow{2}{*}{\textbf{Model}}&
    \multicolumn{2}{c|}{\textbf{Criteo}}&\multicolumn{2}{c|}{\textbf{Avazu}}&\multicolumn{2}{c}{\textbf{MovieLens}}\\
    &AUC&Log Loss& AUC & Log Loss & AUC & Log Loss \\
    \hline\hline
    AutoInt & 0.8138& 0.4380 & 0.7634& 0.3675 & 0.9688 & 0.2202\\
    DS-MLP$_{\text{AutoInt}}$ & 0.8145 & 0.4376& 0.7656 & 0.3664 & 0.9734 & 0.2087 \\
    \hline
    DCNv2 & 0.8132 & 0.4389 & 0.7627 & 0.3677 & 0.9690 & 0.2139 \\
    DS-MLP$_{\text{DCNv2}}$ & 0.8147& 0.4371 & 0.7670 & 0.3657 & 0.9728 & 0.2173\\
    \hline
    xDeepFM & 0.8136 & 0.4382 & 0.7639 & 0.3677 & 0.9685 & 0.2151 \\
    DS-MLP$_{\text{xDeepFM}}$ & 0.8149 &0.4373 & 0.7659 & 0.3664& 0.9714 & 0.2074 \\ 
    \hline
    GDCN & 0.8137 & 0.4386 & 0.7648 & 0.3673 & 0.9685 & 0.2164 \\
    DS-MLP$_{\text{GDCN}}$ & 0.8152 &0.4366 & 0.7662 & 0.3663& 0.9737 & 0.2052 \\ 
    \hline
    FinalMLP & 0.8147 & 0.4374 & 0.7659 & 0.3662 & 0.9716 & 0.2119 \\ 
    DS-MLP$_{\text{FinalMLP}}$ & 0.8148 &0.4431 & 0.7662 & 0.3675& 0.9752 & 0.1971 \\ 
  \bottomrule
\end{tabular}
\end{table}

\subsection{Experimental Analysis}
Furthermore, we conduct a series of experiments to validate the compatibility and scalability of our proposed method. Through these experiments, we can analyze the importance of dual-stream MLP alignment, demonstrate the significance of each component, and examine the effect of hyper-parameter setting.

\subsubsection{Compatibility Analysis}
As previously discussed in Section~\ref{method:dis}, DS-MLP can flexibly adapt to different types of teacher networks in general.  To rigorously validate the compatibility of our approach, we conducted experiments with five prominent and architecturally diverse teacher models: AutoInt, DCNv2, xDeepFM, GDCN, and FinalMLP. This selection was intentionally made to span a broad spectrum of contemporary feature interaction methodologies. Specifically, AutoInt exemplifies attention-based interaction modeling, while DCNv2 and xDeepFM capture explicit high-order interactions at the feature-wise and vector-level granularities, respectively. GDCN introduces gated cross-network structures to enhance interaction learning, and FinalMLP represents MLP-based architectures that employ feature gating for multi-stream fusion. By leveraging this comprehensive set of teacher models, we are able to thoroughly assess whether DS-MLP functions as a genuinely versatile and teacher-agnostic student network.

The experimental results, summarized in Table~\ref{Compatibility Analysis}, clearly demonstrate the effectiveness and robustness of our approach. Across all five distinct teacher architectures, the DS-MLP student model consistently delivers substantial performance improvements over a standard baseline. These findings indicate that the simple yet powerful MLP backbone serves as an effective universal approximator, capable of capturing and internalizing the unique inductive biases inherent in each teacher, such as the selective weighting of attention mechanisms or the structured combinatorial patterns of cross networks. This consistent improvement, achieved without architecture-specific modifications, highlights DS-MLP as a flexible and broadly applicable framework for advancing feature interaction learning.

\subsubsection{Scalability Analysis}
\label{sec:scale}
Scalability is a critical property for real-world industrial applications, where increasing model capacity often leads to improved performance but may also introduce training instability or overfitting risks. 
To evaluate the scalability of our approach, we conducted a scaling experiment designed to compare the performance trajectory of our DS-MLP against a representative specialized model, GDCN. For both models, we systematically enlarged their parameter space by progressively increasing the hidden size of their MLP components. The resulting performance data points were then fitted using a least squares method to derive smooth, interpretable scaling curves, allowing for a clear comparison of their trends.

\begin{figure}[t]
  \centering
  \subcaptionbox{Criteo Scaling}{
    \label{fig:Criteo Scaling }
    \includegraphics[width=0.48\linewidth]{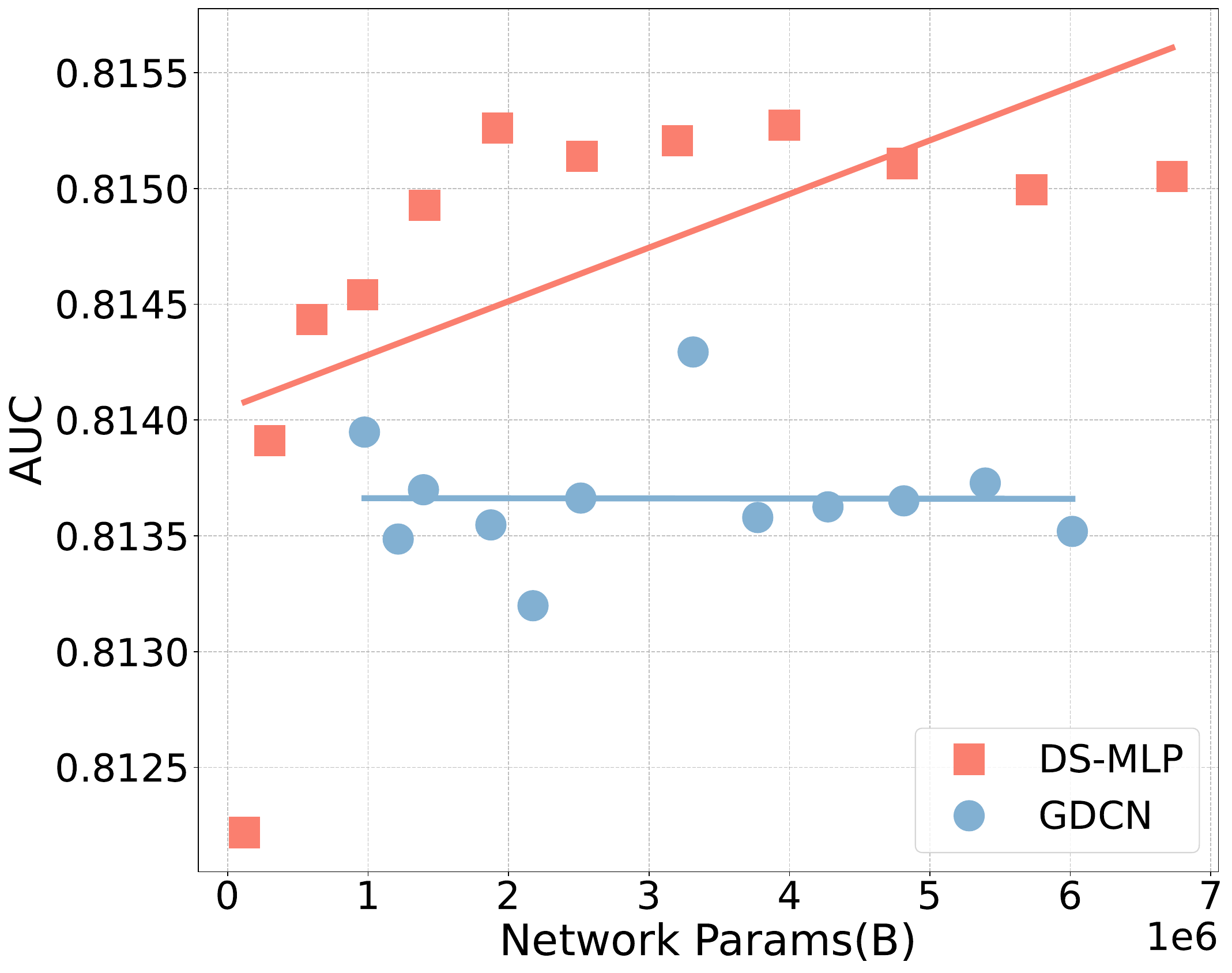}
  }
  \subcaptionbox{Avazu Scaling}{
  \label{fig:Avazu Scaling}
    \includegraphics[width=0.48\linewidth]{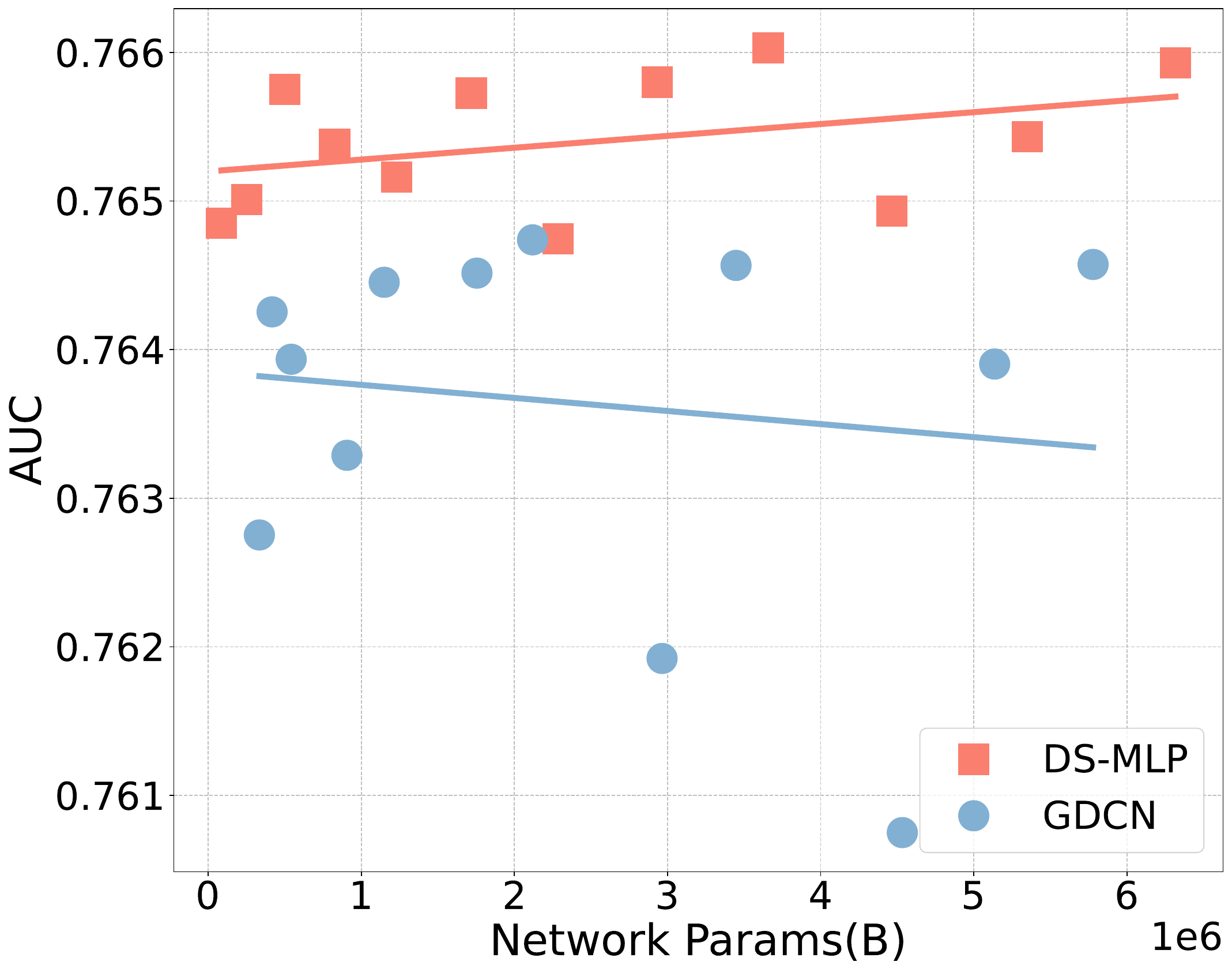}
  }
  \caption{The scaling effect for teacher \emph{v.s.} student. }
  \label{fig:length}
\end{figure}

The results in Figure~\ref{fig:length} reveal a telling divergence in scalability between the two architectures. As the parameter count increases, the performance of the GDCN model saturates and eventually begins to decline, suggesting sensitivity to over-parameterization and potential optimization difficulties. In contrast, our dual-stream MLP exhibits a steady and consistent improvement in performance as its capacity grows, demonstrating robustness to scaling. These observations indicate that while highly specialized architectures may encounter stability or overfitting issues under large capacity regimes, the dual-stream MLP benefits from its simple yet powerful backbone, which is inherently easier to optimize and more resilient to overfitting, making it particularly well-suited for scalable industrial deployment.

\begin{figure}[h]
  \centering
  \includegraphics[width=.7\linewidth]{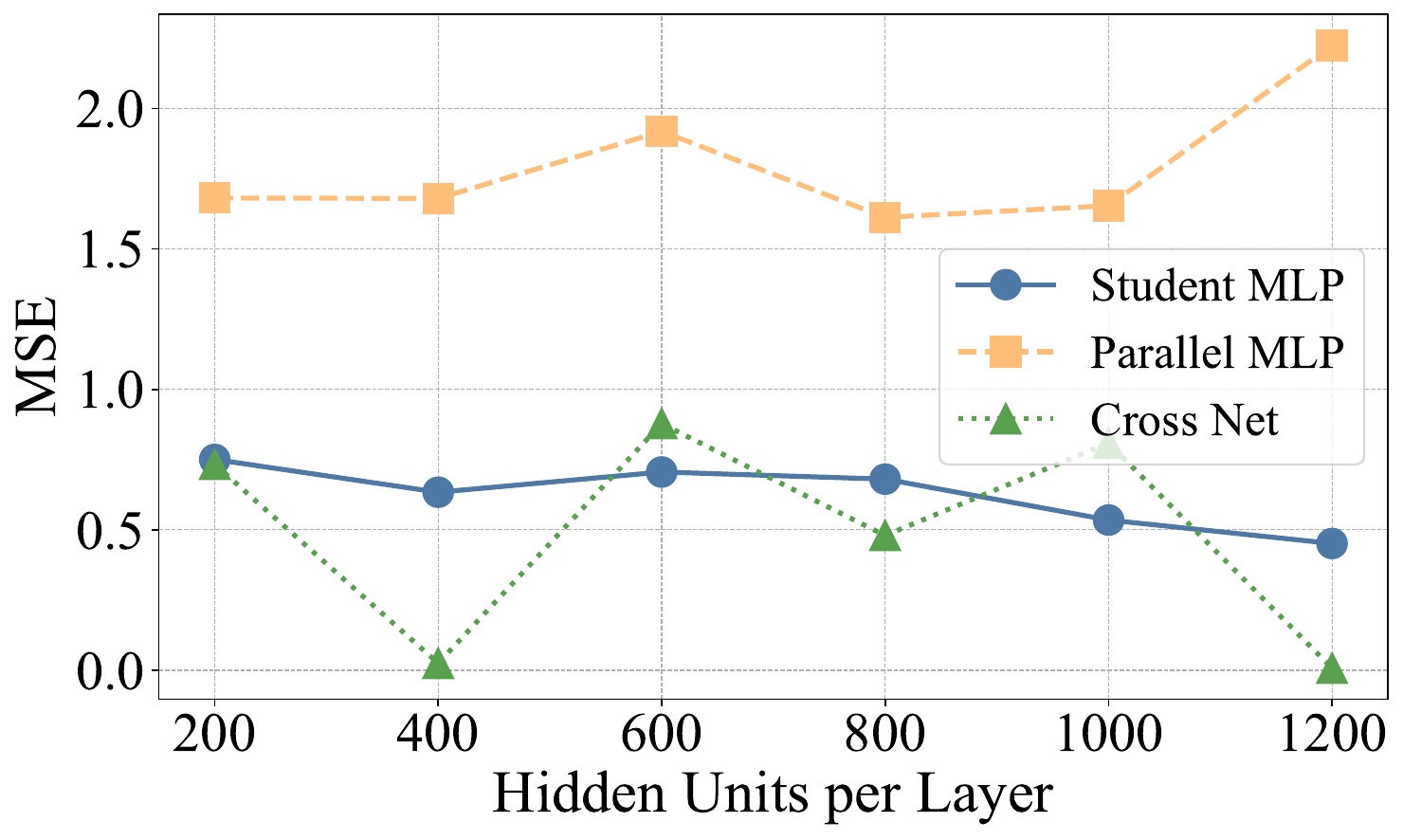}

  \caption{Comparison of MSE on each component in learning explicit feature interactions.}
   \label{fig:mse}
\end{figure}

\subsubsection{Fidelity Analysis of Distilled Explicit Feature Interactions}
In this experiment, we conduct a polynomial fitting study to rigorously evaluate whether the student network is capable of approximating the teacher model’s explicit feature interactions through knowledge distillation. To this end, we first synthesize a dataset comprising three categorical features $(x_1, x_2, x_3)$ and generate corresponding ground-truth labels $y$ using the teacher model (GDCN), thereby constructing a controlled polynomial fitting dataset. To quantify the fidelity of the distillation process, we compute the Mean Squared Error (MSE) between the ground truth and three model components: (i) the CrossNet module of GDCN, which explicitly models feature interactions, (ii) the student MLP within DS-MLP, which is trained to approximate these explicit interactions, and (iii) the parallel MLP of DS-MLP, which focuses on learning implicit interactions and serves as an auxiliary learner.

The results, illustrated in Figure~\ref{fig:mse}, indicate a clear distinction between the student and parallel MLP modules. Across varying MLP sizes, the student MLP consistently achieves lower MSE values and exhibits a convergence pattern closely aligned with the CrossNet module of the teacher model. In contrast, the parallel MLP exhibits a consistently high MSE, confirming that it does not replicate the teacher's logic and remains free to learn complementary, implicit patterns as intended.
These observations confirm that the student module effectively internalizes complex explicit feature interactions through distillation, capturing patterns that are otherwise difficult to learn via implicit modeling alone.

\begin{figure}[t]
  \centering
  \includegraphics[width=.8\linewidth]{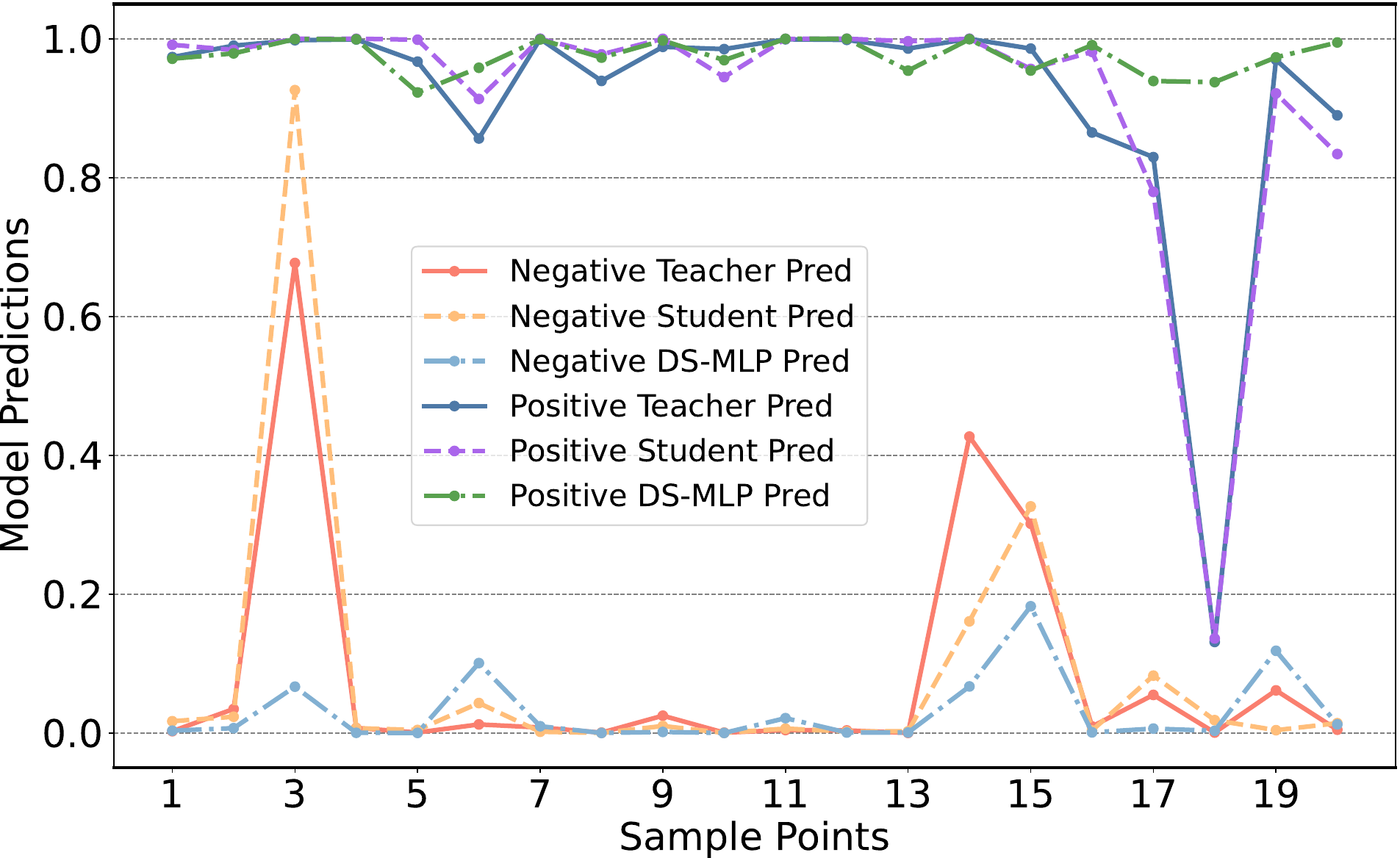}
  \caption{Case study for our approach DS-MLP.}
   \label{fig:case study}
\end{figure}

\subsubsection{Case Study of Dual-Stream Predictions}
\label{exp:case}
In this section, we illustrate the effectiveness of our method and the importance of the dual-stream architecture in Figure~\ref{fig:case study}. 
The $x$-axis represents the sample index, and the $y$-axis represents the prediction score.
We observe that for certain challenging instances, specifically negative sample 3 and positive sample 18, the teacher network produces incorrect predictions, outputting values close to 1 and 0, respectively.
Following the initial distillation phase, the student MLP faithfully mimics its teacher and consequently inherits these same erroneous predictions. This highlights a critical pitfall of standard knowledge distillation, where a student can learn the teacher's biases along with its strengths.

After applying the dual-stream alignment, the parallel MLP actively mitigates the prediction errors generated by the student MLP. This corrective effect exemplifies the unique strength of our dual-stream design: during the alignment phase, the parallel MLP operates independently of the teacher's bias, allowing it to capture complementary decision patterns directly from the data. By providing a corrective signal, it compensates for the mistakes introduced by the distillation process, ultimately enabling DS-MLP to generate correct predictions for these difficult samples. This case study highlights the important role of the dual-stream architecture in enhancing model robustness and predictive accuracy. It shows that combining a distilled student with an auxiliary parallel stream can effectively mitigate teacher biases and improve generalization on challenging prediction tasks.

\subsubsection{Ablation Study}

To gain a deeper understanding of the contribution of each proposed component to the overall performance, we conduct a series of ablation experiments. By systematically removing or replacing individual modules while keeping the remaining parts of the framework unchanged, we are able to isolate their effects on model behaviour. The corresponding results are summarized in Table~\ref{tab:ablation}.  

$\bullet$ \textit{w/o} alignment loss.  
This variant removes the alignment loss defined in Eq.~\eqref{equ:align}, optimizing the model solely with the remaining loss terms. The results reveal a clear and consistent performance drop, underscoring the crucial role of an effective alignment mechanism between the dual-stream MLPs in facilitating the transfer of complementary information. Without such alignment, as discussed in Section~\ref{subsec:align}, the output of one stream tends to dominate, thereby overshadowing the contribution of the other component. Consequently, the two streams fail to cooperate synergistically, leading to a noticeable degradation in the performance of DS-MLP.

\begin{table}[t]
  \caption{The effectiveness of different components in DS-MLP.}\label{tab:ablation}
  \begin{tabular}{c|cc|cc|cc}
    \toprule
    \multirow{2}{*}{\textbf{Model}}&
    \multicolumn{2}{c|}{\textbf{Avazu}}&\multicolumn{2}{c|}{\textbf{Criteo}}&\multicolumn{2}{c}{\textbf{MovieLens}}\\
    &AUC&Log Loss& AUC & Log Loss & AUC & Log Loss \\
    \hline\hline
    DS-MLP & \textbf{0.7670} & \textbf{0.3657} &\textbf{0.8152} & \textbf{0.4366} & \textbf{0.9752} & \textbf{0.1971}\\
    w/o alignment Loss & 0.7641 & 0.3675 & 0.8145 & 0.4376 & 0.9703 & 0.2143 \\
    w/o student MLP & 0.7652 &0.3664 & 0.8137 & 0.4374& 0.9669 & 0.2202 \\ 
    w/o parallel MLP & 0.7648& 0.3672 & 0.8140& 0.4379 & 0.9690 & 0.2191 \\
    w/o dual MLP & 0.7639& 0.3666 & 0.8138 & 0.4382 & 0.9708 & 0.2005\\
    w/o KD & 0.7655 &0.3662 & 0.8135 & 0.4388& 0.9715 & 0.2059 \\ 
  \bottomrule
\end{tabular}
\end{table}

$\bullet$ \textit{w/o} student MLP. This variant bypasses the distillation process entirely and directly aligns and fuses the original teacher model with the parallel MLP. This observation reveals that simply combining the teacher model with an auxiliary MLP is insufficient for effective knowledge transfer. Even though an alignment loss is applied to enforce output consistency, the substantial architectural discrepancy between the highly specialized teacher architecture and the parallel MLP hinders them from forming a coherent and compact representation space. Eliminating the student MLP also sacrifices structural simplification, leading to an inflated parameter count and degraded computational efficiency. Importantly, as demonstrated in Section~\ref{sec:scale}, blindly increasing model parameters does not necessarily yield performance gains. Therefore, the objective of distillation in our framework extends far beyond mere model simplification. Fundamentally, it translates the explicit feature interaction paradigm into a homogeneous structure consistent with the implicit pathway. This structural consistency is precisely what enables optimal alignment between the two streams and endows the overall architecture with robust scaling capabilities.

$\bullet$ \textit{w/o} parallel MLP.
In this variant, only the main MLP is retained for prediction, while the parallel MLP is removed. As a result, the model becomes a single-stream architecture, severely limiting its capacity to capture implicit feature interactions that the teacher-guided main stream might overlook. As demonstrated in Section~\ref{exp:case}, the auxiliary parallel stream plays a key role in mitigating teacher-induced biases and improving generalization on challenging samples. Consequently, removing the parallel MLP leads to a marked degradation in performance, further highlighting its importance in enriching the learned representations and enhancing overall prediction quality.

$\bullet$ \textit{w/o} dual MLP. This variant replaces the dual-stream MLP with the CrossNet structure~\cite{wang2021dcn}.  
The results show that CrossNet fails to effectively absorb and replicate the knowledge of the teacher network through distillation. This phenomenon can be attributed to its highly specialized design, which limits its compatibility with heterogeneous teacher architectures and makes optimization more difficult. By contrast, MLPs are universal function approximators with simple and homogeneous structures, enabling them to mimic diverse teacher behaviors and generalize across a broad range of interaction patterns. This comparison highlights the necessity of adopting an MLP-based backbone as the foundation of our framework, since its flexibility and ease of optimization make it substantially more suitable for knowledge transfer than other highly customized architectures.

$\bullet$ \textit{w/o} KD. 
In this variant, the knowledge distillation stage is removed, resulting in a substantial drop in performance. Trained entirely from scratch without teacher guidance, the dual-stream MLP can only capture implicit feature interactions and fails to inherit the structured knowledge of explicit interactions from the teacher. This finding clearly demonstrates that KD is the cornerstone of our framework: it provides crucial initial guidance and transfers explicit interaction patterns on which the student model can build. Without this foundation, the model struggles to reach optimal performance, underscoring the necessity of jointly modeling explicit and implicit feature interactions for CTR prediction tasks.


Overall, the ablation results offer strong evidence of both the necessity and effectiveness of the proposed components. Each module makes a distinct yet complementary contribution, enabling the model to capture richer feature interactions and to fully leverage teacher guidance. By working in concert, these components allow DS-MLP to consistently achieve superior and more stable performance.


\begin{figure}[t]
  \centering
  \begin{minipage}{0.49\textwidth}
    \centering
    \includegraphics[width=\linewidth]{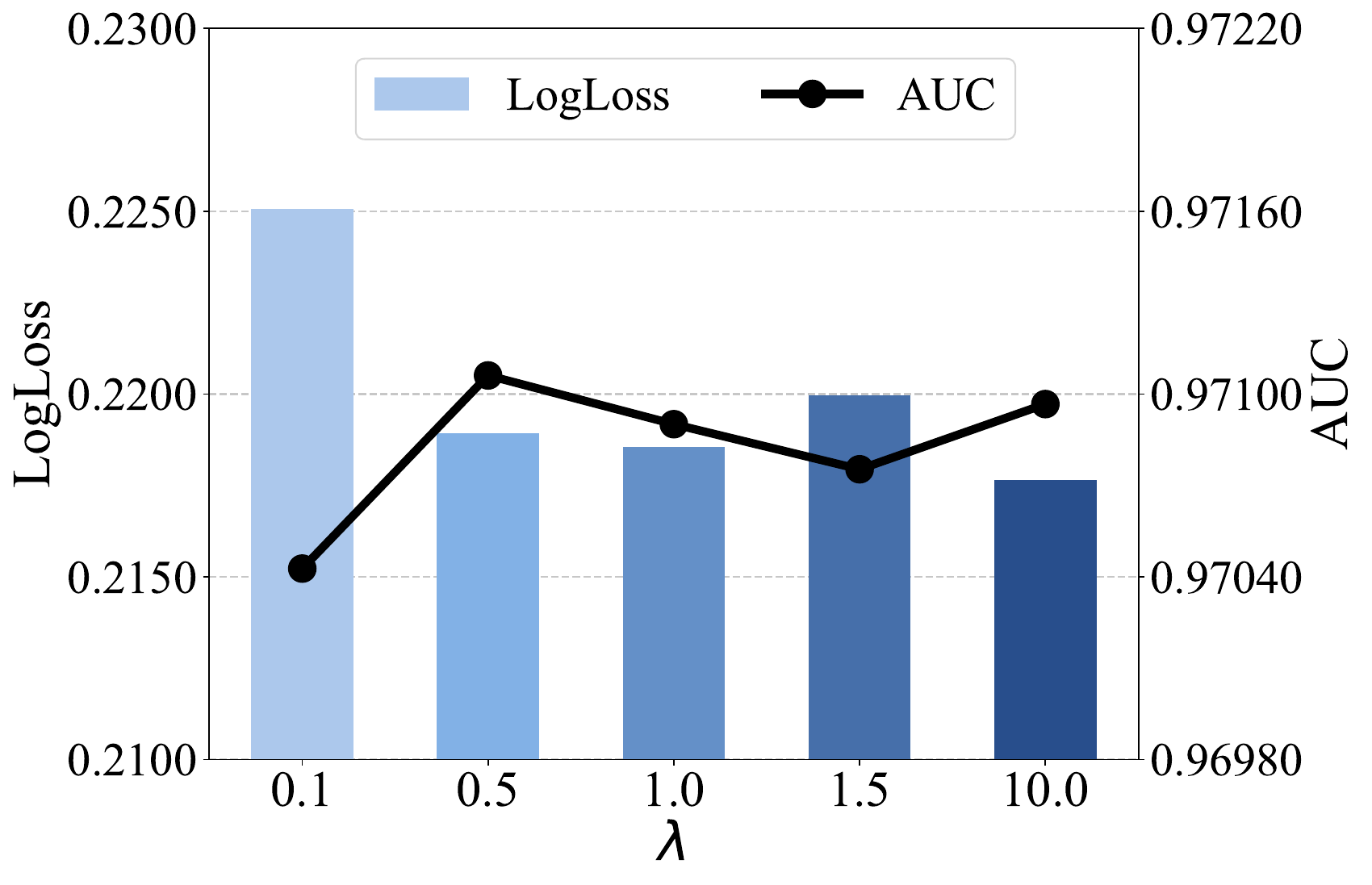}
    \subcaption{Movielens KD Loss}
    \label{fig:Movielens KD Loss}
  \end{minipage}
  \hfill
  \begin{minipage}{0.49\textwidth}
    \centering
    \includegraphics[width=\linewidth]{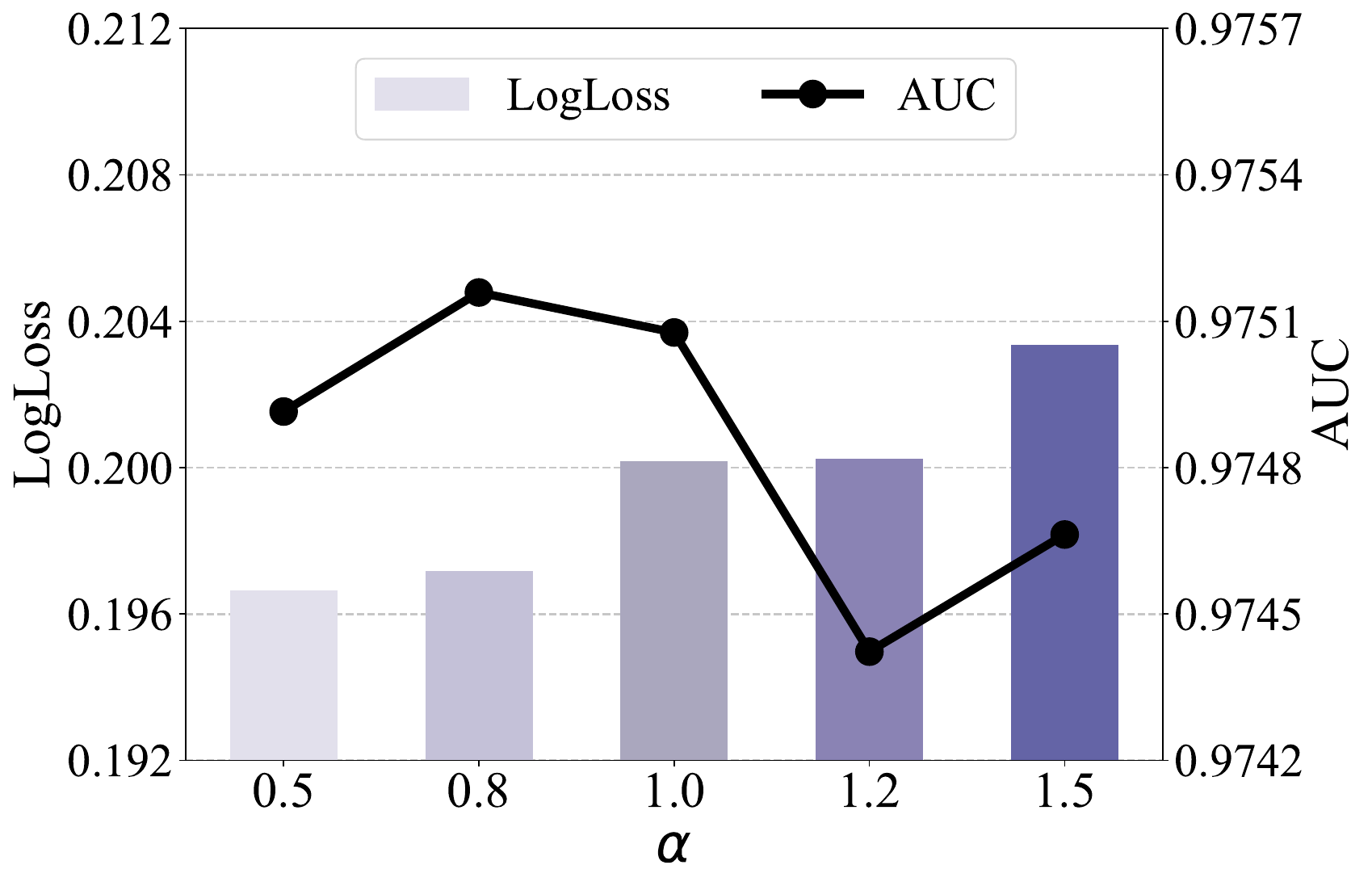}
    \subcaption{Movielens Align Loss}
    \label{fig:Movielens Align Loss}
  \end{minipage}

  \begin{minipage}{0.49\textwidth}
    \centering
    \includegraphics[width=\linewidth]{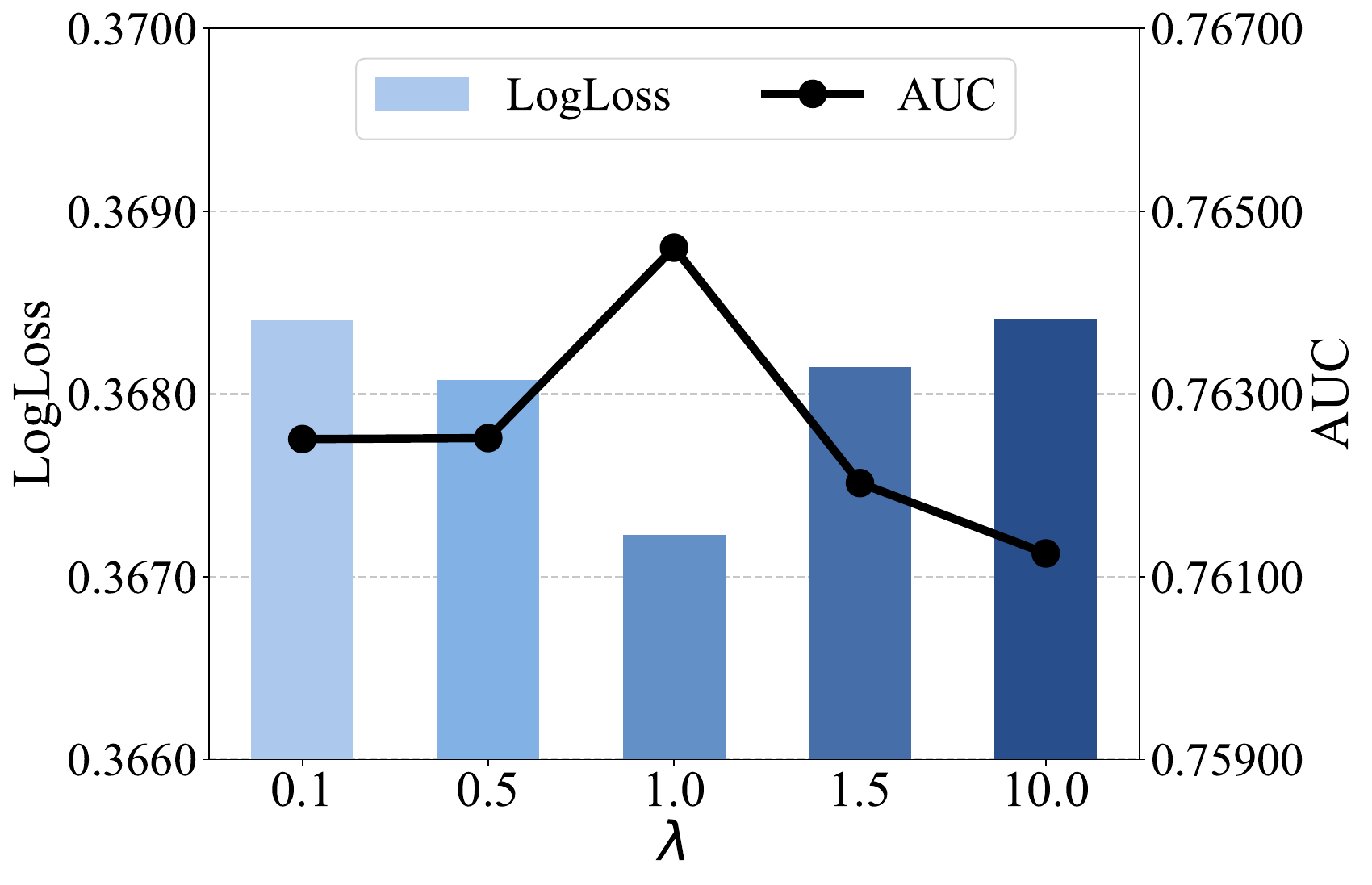}
    \subcaption{Avazu KD Loss}
    \label{fig:Avazu KD loss}
  \end{minipage}
  \hfill
  \begin{minipage}{0.49\textwidth}
    \centering
    \includegraphics[width=\linewidth]{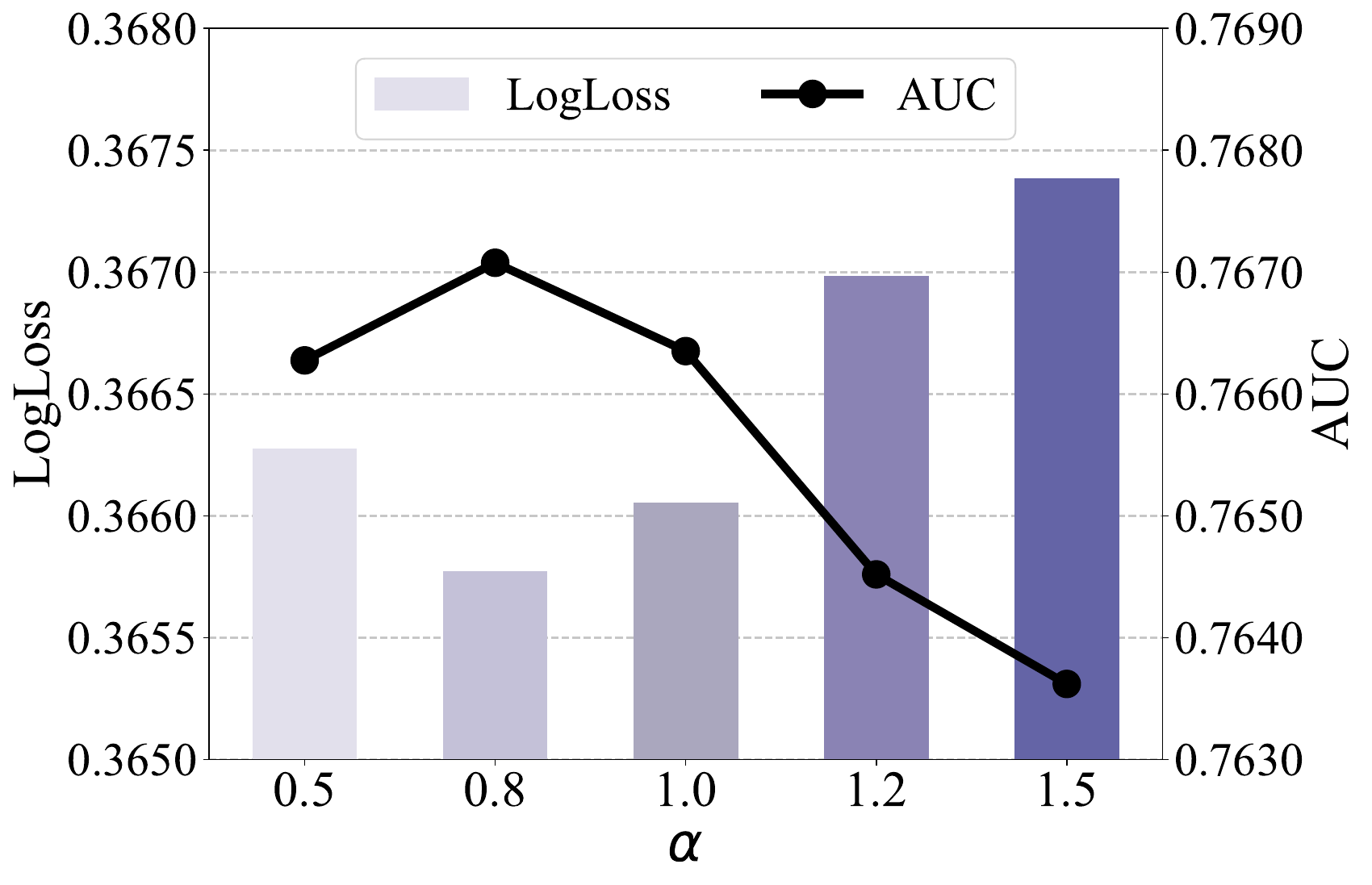}
    \subcaption{Avazu Align Loss}
    \label{fig:Avazu Align Loss}
  \end{minipage}

  \begin{minipage}{0.49\textwidth}
    \centering
    \includegraphics[width=\linewidth]{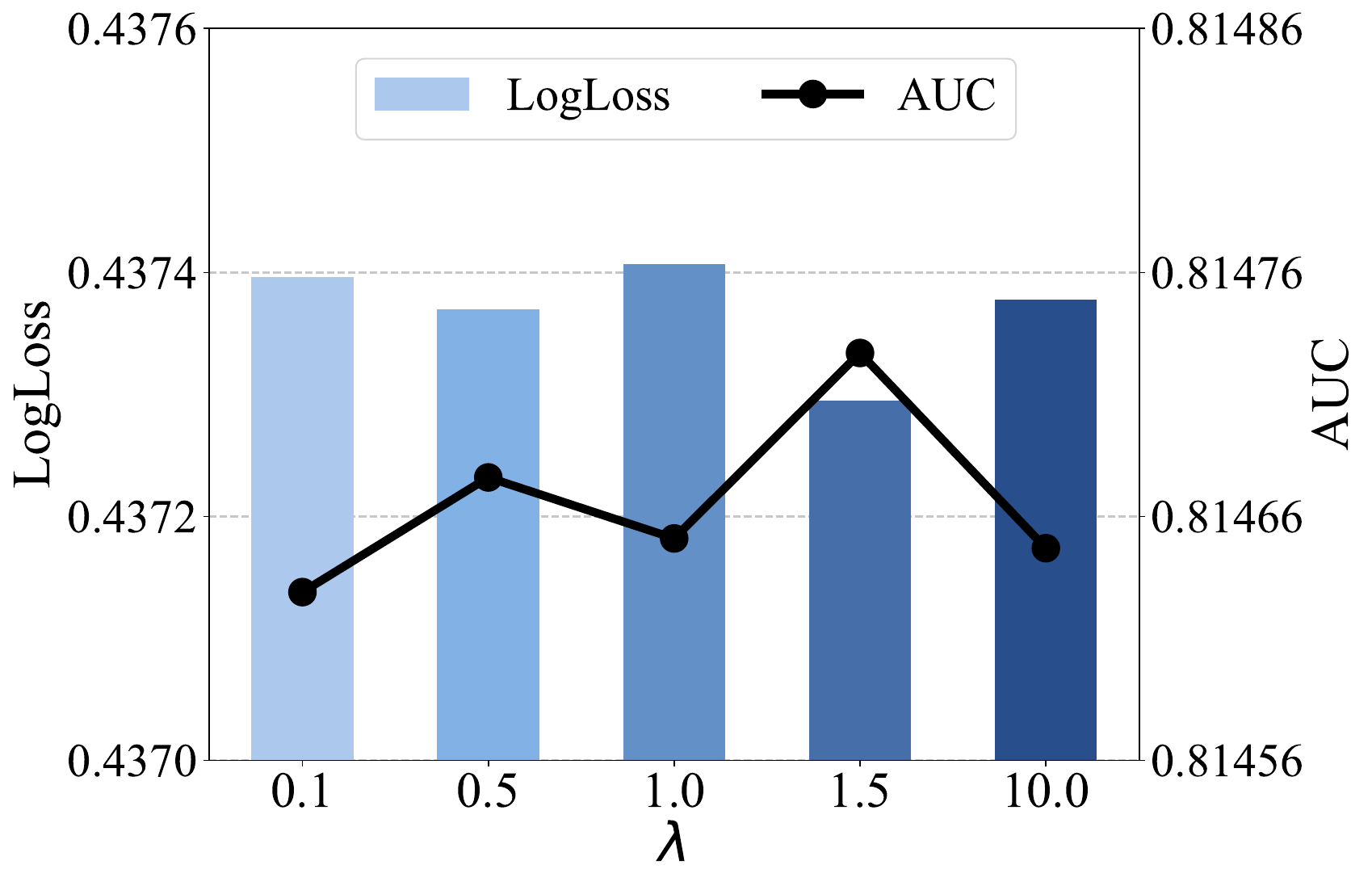}
    \subcaption{Criteo KD Loss}
    \label{fig:Criteo KD Loss}
  \end{minipage}
  \hfill
  \begin{minipage}{0.49\textwidth}
    \centering
    \includegraphics[width=\linewidth]{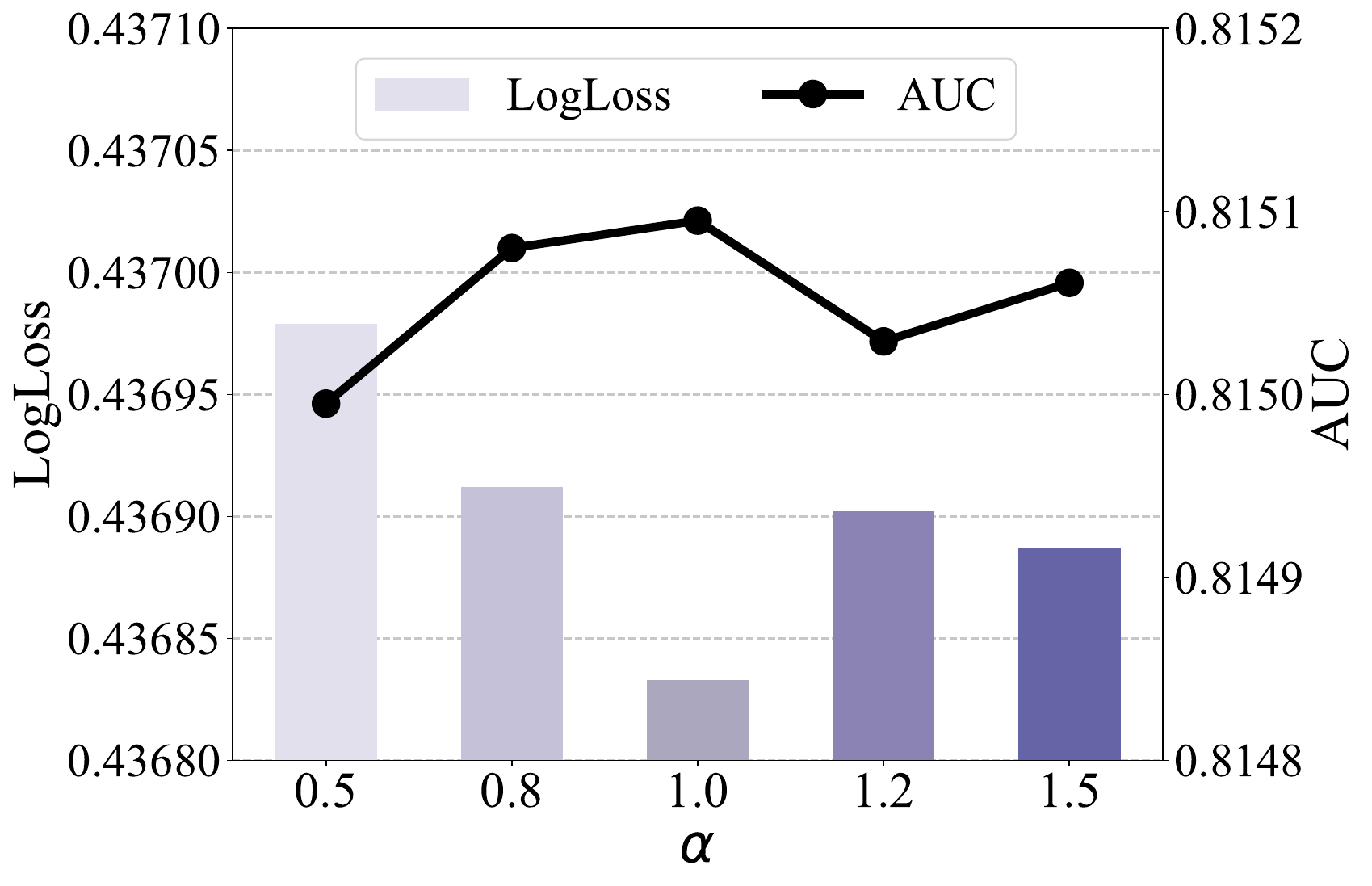}
    \subcaption{Criteo Align Loss}
    \label{fig:Criteo Align Loss}
  \end{minipage}

  \caption{Impact of the coefficients $\lambda$ and $\alpha$ in KD and Align Loss.}
  \label{fig:hyper}
\end{figure}

\subsubsection{Hyperparameter Analysis}
The performance of our framework is influenced by several key hyperparameters. In this section, we conduct an analysis to understand their impact and provide tuning guidance. We vary these hyperparameters and report the resulting performance changes in Figure~\ref{fig:hyper}.


$\bullet$ \emph{Impact of Coefficient $\lambda$}. 
The coefficient $\lambda$ in Eq.~\eqref{equ:KD stage} plays a pivotal role in our framework, as it governs the trade-off between the knowledge distillation loss and the primary CTR prediction loss. To systematically investigate its effect, we vary its value from 0.1 to 10 and report the results in the left panel of Figure~\ref{fig:hyper}. We observe that larger values of $\lambda$ yield more pronounced gains on feature-rich datasets such as Criteo and Avazu. This improvement arises because a stronger weighting on the distillation loss compels the student network to more faithfully inherit the complex, high-order feature interactions already captured by the teacher model. Nevertheless, overly large values of $\lambda$ can lead the student to over-rely on the teacher’s potentially biased soft labels while underutilizing the ground-truth supervision, which may in turn hinder generalization. These results underscore the necessity of selecting an appropriate balance that maximizes knowledge transfer from the teacher while maintaining fidelity to the true data distribution.

$\bullet$ \emph{Impact of Coefficient $\alpha$}.  
In the fine-tuning stage, the parameter $\alpha$ defined in Eq.~\eqref{equ:finetuning} controls the degree of alignment between the two MLP components.
To systematically examine its influence, we vary $\alpha$ from 0.5 to 1.5 while keeping the variation trend of $\beta$ consistent (right panel of Figure~\ref{fig:hyper}). The resulting performance curve peaks at approximately $\alpha=1$, exhibiting an initial rise followed by a decline. This pattern suggests that excessively large values of $\alpha$ overly constrain the two streams, forcing their outputs to become almost indistinguishable and thereby diminishing their ability to learn complementary interaction information. Conversely, too small values of $\alpha$ weaken the alignment effect, preventing the model from effectively reconciling the two representations. The optimal value thus reflects a “sweet spot” where model consistency and representational diversity are well balanced, enabling DS-MLP to leverage both the shared structure and the unique strengths of each stream.

Overall, these findings demonstrate that DS-MLP maintains strong and stable performance across a wide range of hyperparameter choices, indicating its inherent robustness. At the same time, balanced settings of key coefficients such as $\lambda$ and $\alpha$ enhance the integration of teacher guidance and complementary feature interactions, leading to improved performance.

 \section{Related Work}

In this section, we review the related work in the following three aspects, including CTR prediction, two-stream CTR models, and knowledge distillation. 

\subsection{CTR Prediction}
The click-through rate (CTR) prediction task, which aims to estimate the probability of a user clicking an item, is a cornerstone of modern personalized services. The field’s progression has been defined by an ongoing effort to more effectively model the complex feature interactions among user attributes (\eg age, location), item attributes (\eg category, brand), and contextual features (\eg time, device). Early industrial-scale approaches, such as Logistic Regression (LR)~\cite{richardson2007predicting}, were efficient but limited, relying heavily on laborious manual feature engineering. This limitation spurred the development of Factorization Machine (FM)-based methods~\cite{sun2021fm2,pan2018field,xiao2017attentional,juan2016field}, which exploit low-rank factorization to explicitly model second-order feature interactions, particularly in sparse and high-dimensional settings.

The advent of deep learning marked a significant paradigm shift, ushering in an era of diverse and powerful architectures. Initial efforts focused on adapting successful models from other domains to the specific challenges of CTR. For instance, Recurrent Neural Networks (RNNs), such as DSIN~\cite{feng2019deep}, have been employed to capture sequential dependencies in user behavior, while Convolutional Neural Networks (CNNs), such as FGCNN~\cite{liu2019feature} and CFM~\cite{xin2019cfm}, excel at extracting local patterns from feature embeddings. Subsequently, more sophisticated methods emerged to learn relationships between features dynamically. Attention-based networks (\eg AutoInt~\cite{song2019autoint}, GraphFM~\cite{li2021graphfm}, FRNet~\cite{wang2022enhancing}) were introduced to dynamically weight feature importance and better capture global dependencies, while Graph Neural Networks (GNNs), such as FiGNN~\cite{li2019fi} and DG\_ENN~\cite{guo2021dual}, further extended modeling capabilities to complex, structured entity relationships in user–item graphs.
More recent research has pushed towards even greater flexibility and automation. A notable trend is the development of adaptive interaction mechanisms, such as RFM~\cite{tian2024rotative} and EulerNet~\cite{tian2023eulernet}. These models apply advanced mathematical transformations, including operations in complex vector spaces, to flexibly learn feature interactions in an order-agnostic manner, effectively mitigating the underfitting and overfitting risks associated with fixed-order designs.
As the manual design of these intricate architectures has become a bottleneck, the field has also turned to AutoML-based approaches (\eg AIM~\cite{zhu2021aim}, Autofeature~\cite{khawar2020autofeature}, Amer~\cite{zhao2020amer}), which automate feature learning and model selection to achieve competitive performance with minimal human intervention.

More recently, the field has begun exploring new frontiers beyond architectural diversity, focusing on scalability and paradigm transformation. 
On one hand, inspired by the success of scaling laws~\cite{kaplan2020scaling} in large language models, industrial research~\cite{hou2026kunlun,guan2025make,lai2025exploring} has increasingly focused on systematic scaling strategies for recommendation systems. 
For example, WuKong~\cite{zhang2024wukong} and RankMixer~\cite{zhu2025rankmixer} show that scaling laws in recommendation can be realized through structured capacity expansion with stacked factorization machines and large-scale mixture-based ranking architectures, alleviating the capacity limits of traditional models.
On the other hand, the emergence of generative recommendation~\cite{rajput2023recommender,deng2025onerec} has motivated a paradigm shift from discriminative feature interaction to generative modeling. 
Rather than directly predicting clicks based on handcrafted interaction functions, recent approaches~\cite{kong2025generative,zhang2025dgenctr} attempt to capture feature dependencies through generative processes. For instance, SFG~\cite{yin2025feature} reformulates CTR prediction as supervised feature generation via an encoder–decoder framework, while GenCI~\cite{ou2026genci} employs generative intent modeling to capture dynamic user preferences.

This evolutionary trajectory, progressing from manual engineering to diverse deep architectures, scalable industrial designs, and now paradigm-level innovations, highlights the continuous pursuit of more powerful and intelligent solutions for modeling feature interactions.

\subsection{Two-Stream CTR Models}
To better capture complex feature interactions, modern CTR prediction has increasingly adopted dual-stream architectures~\cite{guo2017deepfm,wang2025dlf,wang2017deep}, which extract complementary information from two perspectives: implicit and explicit.  Implicit interactions refer to cross-feature relationships that are not predefined but instead emerge indirectly through nonlinear transformations. 
Multi-Layer Perceptrons (MLPs~\cite{popescu2009multilayer}) are well suited for this purpose. They act as universal function approximators. By stacking nonlinear layers, MLPs automatically combine features and build complex high-order dependencies from data without manual feature engineering.
However, the behavior of these implicit mechanisms is inherently opaque: it is difficult to ascertain which specific interactions have actually been learned, or whether the learned cross terms are meaningful at all, thus limiting model interpretability and diagnostics.

In contrast to implicit models, the explicit stream is designed to model feature interactions in a structured and interpretable manner. This is typically achieved using well-defined mathematical operations such as the inner product, outer product, or element-wise product to directly simulate feature crosses. Foundational models such as FM~\cite{rendle2010factorization}, PNN~\cite{qu2018product}, and CrossNet~\cite{wang2017deep} pioneered this approach. Extensive research~\cite{sun2021fm2,pan2018field,juan2016field} has highlighted the critical role of second-order feature interactions in CTR prediction, showing that modeling pairwise dependencies between features can significantly improve predictive performance. Nevertheless, methods that focus solely on second-order interactions are inherently limited in their ability to capture higher-order dependencies, which has motivated recent studies~\cite{wang2021masknet,yu2020deep,lu2021dual,cai2021arm,tian2023eulernet} to develop more sophisticated architectures capable of explicitly modeling complex, higher-order feature interactions. For instance, DCNv2~\cite{wang2021dcn} and xDeepFM~\cite{lian2018xdeepfm} stack multiple layers of CrossNet and CIN to model high-order explicit interactions, while AutoInt employs a multi-head self-attention mechanism to adaptively capture higher-order terms.
Final~\cite{zhu2023final} rapidly models high-order feature interactions through its exponentially growing structure, with self-distillation enabling effective fusion across modules. FinalMLP~\cite{mao2023finalmlp} further combines MLPs with feature gating and interaction aggregation to form a dual-stream architecture capturing both implicit and explicit interactions. 

Despite these advances, most existing models, although well designed and structurally innovative, suffer from substantial architectural complexity, which raises computational cost and hinders deployment in real-world systems. Furthermore, the implicit and explicit components often exhibit highly heterogeneous structures, and directly aggregating their outputs can lead to inconsistencies and suboptimal predictions. This highlights the necessity of alignment mechanisms to harmonize the two streams. In our approach, both the implicit and explicit branches share the same MLP backbone, which ensures structural compatibility of their outputs and facilitates a more effective and seamless alignment between the streams.

\subsection{Knowledge Distillation} 
Knowledge distillation~\cite{hinton2015distilling} aims to transfer knowledge from a teacher model to a student network through the teacher model's outputs as supervisory signals. Beyond the conventional soft label~\cite{hinton2015distilling} and hint regression~\cite{FitNet} strategies, subsequent studies have broadened the scope of knowledge transfer to include inter-layer flow~\cite{FitNet}, cross-sample similarity~\cite{KDGift}, attention-based mechanisms~\cite{attention}, and other advanced techniques. These approaches allow student networks to approximate the performance of their teacher counterparts while substantially reducing computational costs, thereby improving their practicality for real-world deployment. More recently, ensemble distillation has been proposed to aggregate knowledge from multiple teacher models, achieving strong results across diverse tasks such as image classification~\cite{KD-translation} and machine translation~\cite{MEAL}. The success of these methods across various domains has also stimulated growing interest within the recommendation systems community. In large-scale recommendation scenarios, where high-dimensional sparse features and intricate feature interactions are common, distillation provides an effective means of balancing predictive performance and computational efficiency. By narrowing the performance gap between large and small models, KD-based approaches~\cite{xu2020privileged,zhu2020ensembled,tian2023directed,zhou2018rocket} have emerged as a particularly promising direction for recommender systems.

Existing knowledge distillation approaches exhibit considerable diversity in the mechanisms by which knowledge is transferred from teacher to student. Some methods concentrate primarily on aligning the teacher’s output logits, whereas others emphasize the matching of intermediate hidden representations to provide richer supervisory signals for the student network. For instance, ECKD~\cite{zhu2020ensembled} aggregates knowledge from multiple teacher networks to exploit their complementary strengths, yet its effectiveness is highly dependent on the quality and diversity of the teachers, and excessive reliance on teacher signals can impede the student’s independent learning. To alleviate the accuracy loss induced during distillation, DAGFM~\cite{tian2023directed} introduces a directed acyclic graph factorization machine to better capture complex feature relationships and narrow the gap between teacher and student. Despite these advances, KD-based methods continue to face critical challenges, including teacher–student mismatch, insufficient joint modeling of explicit and implicit interactions, and overfitting to the teacher distribution, all of which collectively undermine the generalization capacity of the student model. This highlights the necessity of designing more effective distillation and post-distillation procedures. In response, we propose an alignment fine-tuning stage that directly addresses teacher–student mismatch and strengthens the unified modeling of explicit and implicit feature interactions, thereby enabling the student model to achieve stronger robustness and enhanced generalization.

\section{Conclusion}

In this paper, we introduced the Dual-Stream MLP (DS-MLP), a framework for CTR prediction that is simple, efficient, and highly capable. By leveraging the universal approximation property of MLPs, our framework effectively distills structured knowledge from high-capacity teacher models that capture complex and high-order explicit feature interactions. This process builds a strong and transferable foundation of generalized knowledge for the student model, while preserving the efficiency of lightweight architectures.

To further enhance representational capacity, we incorporate a parallel auxiliary MLP stream dedicated to modeling implicit feature interactions that may be underrepresented in the main stream. Since the two streams are designed to capture complementary explicit and implicit interaction patterns, DS-MLP introduces two alignment strategies operating at both the hidden-state and prediction levels. Batch-wise regularization stabilizes the outputs of both streams and brings them to a comparable numeric range. Meanwhile, direct task supervision ensures that each stream is aware of the target objective, functions as an independent predictor, and thus adjusts its outputs to achieve mutual compatibility. This enhances the task-specific capacity of each stream while preventing the stronger component from overwhelming the weaker one. Through the above optimization scheme, DS-MLP achieves a seamless and structurally consistent integration of explicit and implicit interaction modeling, resulting in more robust and interpretable feature representations.

Extensive experiments on multiple public benchmarks show that DS-MLP consistently outperforms strong baselines and achieves state-of-the-art performance. Importantly, its reliance on a pure MLP backbone ensures remarkable computational efficiency and low inference latency, making it both practical and scalable for real-world recommendation systems.

For future work, we identify several promising directions to further enhance the DS-MLP framework.
First, we aim to extend the dual-stream architecture to a multi-stream design, potentially leveraging a Mixture-of-Experts (MoE) paradigm to capture a broader spectrum of feature interaction patterns. Second, we plan to investigate adaptive strategies that jointly determine the optimal depth, width, and parameter-sharing of MLP components while dynamically balancing the contributions of explicit and implicit streams during training. These extensions are expected to enhance representational expressiveness, improve prediction robustness, and allow the model to better manage computational efficiency and hardware constraints.

\begin{acks}
This paper was partially supported by the National Natural Science Foundation of China No. 92470205 and Beijing Major Science and Technology Project under Contract no. Z251100008425002. Xin Zhao is the corresponding author.
\end{acks}

\bibliographystyle{ACM-Reference-Format}
\bibliography{main}

@String{Computing = "Computing" }

@String{Computer = "{IEEE} Computer" }

@String{Springer = "Springer-Verlag" }

@article{popescu2009multilayer,
  title={Multilayer perceptron and neural networks},
  author={Popescu, Marius-Constantin and Balas, Valentina E and Perescu-Popescu, Liliana and Mastorakis, Nikos},
  journal={WSEAS Transactions on Circuits and Systems},
  volume={8},
  number={7},
  pages={579--588},
  year={2009}
}

@inproceedings{wang2025dlf,
  title={DLF: Enhancing Explicit-Implicit Interaction via Dynamic Low-Order-Aware Fusion for CTR Prediction},
  author={Wang, Kefan and Wang, Hao and Guo, Wei and Liu, Yong and Lin, Jianghao and Lian, Defu and Chen, Enhong},
  booktitle={Proceedings of the 48th International ACM SIGIR Conference on Research and Development in Information Retrieval},
  pages={2213--2223},
  year={2025}
}

@ArtifactSoftware{R,
    title = {R: A Language and Environment for Statistical Computing},
    author = {{R Core Team}},
    organization = {R Foundation for Statistical Computing},
    address = {Vienna, Austria},
    year = {2019},
    url = {https://www.R-project.org/},
}

@inproceedings{richardson2007predicting,
 title={Predicting clicks: estimating the click-through rate for new ads},
 author={Richardson, Matthew and Dominowska, Ewa and Ragno, Robert},
 booktitle={Proceedings of the 16th international conference on World Wide Web},
 pages={521--530},
 year={2007}
}

@inproceedings{wang2021dcn,
  title={DCN V2: Improved deep \& cross network and practical lessons for web-scale learning to rank systems},
  author={Wang, Ruoxi and Shivanna, Rakesh and Cheng, Derek and Jain, Sagar and Lin, Dong and Hong, Lichan and Chi, Ed},
  booktitle={Proceedings of the Web Conference 2021},
  pages={1785--1797},
  year={2021}
}

@incollection{wang2017deep,
  title={Deep \& cross network for ad click predictions},
  author={Wang, Ruoxi and Fu, Bin and Fu, Gang and Wang, Mingliang},
  booktitle={Proceedings of the ADKDD'17},
  pages={1--7},
  year={2017}
}

@inproceedings{rendle2010factorization,
  title={Factorization machines},
  author={Rendle, Steffen},
  booktitle={2010 IEEE International conference on data mining},
  pages={995--1000},
  year={2010},
  organization={IEEE}
}

@inproceedings{pan2018field,
  title={Field-weighted factorization machines for click-through rate prediction in display advertising},
  author={Pan, Junwei and Xu, Jian and Ruiz, Alfonso Lobos and Zhao, Wenliang and Pan, Shengjun and Sun, Yu and Lu, Quan},
  booktitle={Proceedings of the 2018 World Wide Web Conference},
  pages={1349--1357},
  year={2018}
}

@article{pinkus1999approximation,
  title={Approximation theory of the MLP model in neural networks},
  author={Pinkus, Allan},
  journal={Acta numerica},
  volume={8},
  pages={143--195},
  year={1999},
  publisher={Cambridge University Press}
}

@inproceedings{du2019gradient,
  title={Gradient descent finds global minima of deep neural networks},
  author={Du, Simon and Lee, Jason and Li, Haochuan and Wang, Liwei and Zhai, Xiyu},
  booktitle={International conference on machine learning},
  pages={1675--1685},
  year={2019},
  organization={PMLR}
}

@article{hornik1989multilayer,
  title={Multilayer feedforward networks are universal approximators},
  author={Hornik, Kurt and Stinchcombe, Maxwell and White, Halbert},
  journal={Neural networks},
  volume={2},
  number={5},
  pages={359--366},
  year={1989},
  publisher={Elsevier}
}

@article{zhang2025dgenctr,
  title={DGenCTR: Towards a Universal Generative Paradigm for Click-Through Rate Prediction via Discrete Diffusion},
  author={Zhang, Moyu and Chen, Yun and Jin, Yujun and Hu, Jinxin and Zhang, Yu},
  journal={arXiv preprint arXiv:2508.14500},
  year={2025}
}

@article{kong2025generative,
  title={Generative Click-through Rate Prediction with Applications to Search Advertising},
  author={Kong, Lingwei and Wang, Lu and Peng, Changping and Lin, Zhangang and Law, Ching and Shao, Jingping},
  journal={arXiv preprint arXiv:2507.11246},
  year={2025}
}

@inproceedings{lai2025exploring,
  title={Exploring Scaling Laws of CTR Model for Online Performance Improvement},
  author={Lai, Weijiang and Jin, Beihong and Zhang, Jiongyan and Zheng, Yiyuan and Dong, Jian and Cheng, Jia and Lei, Jun and Wang, Xingxing},
  booktitle={Proceedings of the Nineteenth ACM Conference on Recommender Systems},
  pages={114--123},
  year={2025}
}

@article{guan2025make,
  title={Make It Long, Keep It Fast: End-to-End 10k-Sequence Modeling at Billion Scale on Douyin},
  author={Guan, Lin and Yang, Jia-Qi and Zhao, Zhishan and Zhang, Beichuan and Sun, Bo and Luo, Xuanyuan and Ni, Jinan and Li, Xiaowen and Qi, Yuhang and Fan, Zhifang and others},
  journal={arXiv preprint arXiv:2511.06077},
  year={2025}
}

@article{li2024fcn,
  title={FCN: Fusing Exponential and Linear Cross Network for Click-Through Rate Prediction},
  author={Li, Honghao and Zhang, Yiwen and Zhang, Yi and Li, Hanwei and Sang, Lei and Zhu, Jieming},
  journal={arXiv preprint arXiv:2407.13349},
  year={2024}
}

@article{zhang2024wukong,
  title={Wukong: Towards a scaling law for large-scale recommendation},
  author={Zhang, Buyun and Luo, Liang and Chen, Yuxin and Nie, Jade and Liu, Xi and Guo, Daifeng and Zhao, Yanli and Li, Shen and Hao, Yuchen and Yao, Yantao and others},
  journal={arXiv preprint arXiv:2403.02545},
  year={2024}
}

@article{blondel2016higher,
  title={Higher-order factorization machines},
  author={Blondel, Mathieu and Fujino, Akinori and Ueda, Naonori and Ishihata, Masakazu},
  journal={Advances in Neural Information Processing Systems},
  volume={29},
  year={2016}
}

@inproceedings{xin2019cfm,
  title={CFM: Convolutional factorization machines for context-aware recommendation.},
  author={Xin, Xin and Chen, Bo and He, Xiangnan and Wang, Dong and Ding, Yue and Jose, Joemon M},
  booktitle={IJCAI},
  volume={19},
  pages={3926--3932},
  year={2019}
}

@inproceedings{chen2021enhancing,
  title={Enhancing explicit and implicit feature interactions via information sharing for parallel deep CTR models},
  author={Chen, Bo and Wang, Yichao and Liu, Zhirong and Tang, Ruiming and Guo, Wei and Zheng, Hongkun and Yao, Weiwei and Zhang, Muyu and He, Xiuqiang},
  booktitle={Proceedings of the 30th ACM international conference on information \& knowledge management},
  pages={3757--3766},
  year={2021}
}

@inproceedings{lian2018xdeepfm,
  title={xdeepfm: Combining explicit and implicit feature interactions for recommender systems},
  author={Lian, Jianxun and Zhou, Xiaohuan and Zhang, Fuzheng and Chen, Zhongxia and Xie, Xing and Sun, Guangzhong},
  booktitle={Proceedings of the 24th ACM SIGKDD international conference on knowledge discovery \& data mining},
  pages={1754--1763},
  year={2018}
}

@article{guo2017deepfm,
  title={DeepFM: a factorization-machine based neural network for CTR prediction},
  author={Guo, Huifeng and Tang, Ruiming and Ye, Yunming and Li, Zhenguo and He, Xiuqiang},
  journal={arXiv preprint arXiv:1703.04247},
  year={2017}
}

@inproceedings{song2019autoint,
  title={Autoint: Automatic feature interaction learning via self-attentive neural networks},
  author={Song, Weiping and Shi, Chence and Xiao, Zhiping and Duan, Zhijian and Xu, Yewen and Zhang, Ming and Tang, Jian},
  booktitle={Proceedings of the 28th ACM International Conference on Information and Knowledge Management},
  pages={1161--1170},
  year={2019}
}

@inproceedings{tian2023directed,
  title={Directed acyclic graph factorization machines for CTR prediction via knowledge distillation},
  author={Tian, Zhen and Bai, Ting and Zhang, Zibin and Xu, Zhiyuan and Lin, Kangyi and Wen, Ji-Rong and Zhao, Wayne Xin},
  booktitle={Proceedings of the Sixteenth ACM International Conference on Web Search and Data Mining},
  pages={715--723},
  year={2023}
}

@inproceedings{huang2019fibinet,
  title={FiBiNET: combining feature importance and bilinear feature interaction for click-through rate prediction},
  author={Huang, Tongwen and Zhang, Zhiqi and Zhang, Junlin},
  booktitle={Proceedings of the 13th ACM Conference on Recommender Systems},
  pages={169--177},
  year={2019}
}

@inproceedings{sun2021fm2,
  title={Fm2: Field-matrixed factorization machines for recommender systems},
  author={Sun, Yang and Pan, Junwei and Zhang, Alex and Flores, Aaron},
  booktitle={Proceedings of the Web Conference 2021},
  pages={2828--2837},
  year={2021}
}

@inproceedings{juan2016field,
  title={Field-aware factorization machines for CTR prediction},
  author={Juan, Yuchin and Zhuang, Yong and Chin, Wei-Sheng and Lin, Chih-Jen},
  booktitle={Proceedings of the 10th ACM conference on recommender systems},
  pages={43--50},
  year={2016}
}

@article{deng2025onerec,
  title={Onerec: Unifying retrieve and rank with generative recommender and iterative preference alignment},
  author={Deng, Jiaxin and Wang, Shiyao and Cai, Kuo and Ren, Lejian and Hu, Qigen and Ding, Weifeng and Luo, Qiang and Zhou, Guorui},
  journal={arXiv preprint arXiv:2502.18965},
  year={2025}
}

@article{hou2026kunlun,
  title={Kunlun: Establishing Scaling Laws for Massive-Scale Recommendation Systems through Unified Architecture Design},
  author={Hou, Bojian and Liu, Xiaolong and Liu, Xiaoyi and Xu, Jiaqi and Badr, Yasmine and Hang, Mengyue and Chanpuriya, Sudhanshu and Zhou, Junqing and Yang, Yuhang and Xu, Han and others},
  journal={arXiv preprint arXiv:2602.10016},
  year={2026}
}

@article{kaplan2020scaling,
  title={Scaling laws for neural language models},
  author={Kaplan, Jared and McCandlish, Sam and Henighan, Tom and Brown, Tom B and Chess, Benjamin and Child, Rewon and Gray, Scott and Radford, Alec and Wu, Jeffrey and Amodei, Dario},
  journal={arXiv preprint arXiv:2001.08361},
  year={2020}
}

@article{rajput2023recommender,
  title={Recommender systems with generative retrieval},
  author={Rajput, Shashank and Mehta, Nikhil and Singh, Anima and Hulikal Keshavan, Raghunandan and Vu, Trung and Heldt, Lukasz and Hong, Lichan and Tay, Yi and Tran, Vinh and Samost, Jonah and others},
  journal={Advances in Neural Information Processing Systems},
  volume={36},
  pages={10299--10315},
  year={2023}
}

@article{ou2026genci,
  title={GenCI: Generative Modeling of User Interest Shift via Cohort-based Intent Learning for CTR Prediction},
  author={Ou, Kesha and Tian, Zhen and Zhao, Wayne Xin and Lu, Hongyu and Wen, Ji-Rong},
  journal={arXiv preprint arXiv:2601.18251},
  year={2026}
}

@inproceedings{zhu2025rankmixer,
  title={Rankmixer: Scaling up ranking models in industrial recommenders},
  author={Zhu, Jie and Fan, Zhifang and Zhu, Xiaoxie and Jiang, Yuchen and Wang, Hangyu and Han, Xintian and Ding, Haoran and Wang, Xinmin and Zhao, Wenlin and Gong, Zhen and others},
  booktitle={Proceedings of the 34th ACM International Conference on Information and Knowledge Management},
  pages={6309--6316},
  year={2025}
}

@article{yin2025feature,
  author       = {Mingjia Yin and
                  Junwei Pan and
                  Hao Wang and
                  Ximei Wang and
                  Shangyu Zhang and
                  Jie Jiang and
                  Defu Lian and
                  Enhong Chen},
  title        = {From Feature Interaction to Feature Generation: {A} Generative Paradigm
                  of {CTR} Prediction Models},
  booktitle    = {{ICML}},
  series       = {Proceedings of Machine Learning Research},
  volume       = {267},
  publisher    = {{PMLR} / OpenReview.net},
  year         = {2025}
}

@article{hinton2015distilling,
  title={Distilling the knowledge in a neural network},
  author={Hinton, Geoffrey and Vinyals, Oriol and Dean, Jeff},
  journal={arXiv preprint arXiv:1503.02531},
  year={2015}
}

@inproceedings{FitNet,
  author    = {Adriana Romero and
               Nicolas Ballas and
               Samira Ebrahimi Kahou and
               Antoine Chassang and
               Carlo Gatta and
               Yoshua Bengio},
  title     = {FitNets: Hints for Thin Deep Nets},
  booktitle = {Proceedings of International Conference on Learning Representations, {(ICLR)}},
  year      = {2015},
}

@inproceedings{KDGift,
  author    = {Junho Yim and
               Donggyu Joo and
               Jihoon Bae and
               Junmo Kim},
  title     = {A Gift from Knowledge Distillation: Fast Optimization, Network Minimization
               and Transfer Learning},
  booktitle = {IEEE Conference on Computer Vision and Pattern Recognition (CVPR)},
  year      = {2017}
}

@inproceedings{attention,
  author    = {Sergey Zagoruyko and
               Nikos Komodakis},
  title     = {Paying More Attention to Attention: Improving the Performance of Convolutional
               Neural Networks via Attention Transfer},
  booktitle = {International Conference on Learning Representations, {(ICLR)} 2017},
  year      = {2017}
}

@inproceedings{KD-translation,
  author    = {Xu Tan and
               Yi Ren and
               Di He and
               Tao Qin and
               Zhou Zhao and
               Tie{-}Yan Liu},
  title     = {Multilingual Neural Machine Translation with Knowledge Distillation},
  booktitle = {7th International Conference on Learning Representations ({ICLR})},
  year      = {2019},
}

@inproceedings{MEAL,
  author    = {Zhiqiang Shen and
               Zhankui He and
               Xiangyang Xue},
  title     = {{MEAL:} Multi-Model Ensemble via Adversarial Learning},
  booktitle = {The Thirty-Third {AAAI} Conference on Artificial Intelligence ({AAAI})},
  pages     = {4886--4893},
  year      = {2019}
}

@article{qu2018product,
  title={Product-based neural networks for user response prediction over multi-field categorical data},
  author={Qu, Yanru and Fang, Bohui and Zhang, Weinan and Tang, Ruiming and Niu, Minzhe and Guo, Huifeng and Yu, Yong and He, Xiuqiang},
  journal={ACM Transactions on Information Systems (TOIS)},
  volume={37},
  number={1},
  pages={1--35},
  year={2018},
  publisher={ACM New York, NY, USA}
}

@inproceedings{cheng2016wide,
  title={Wide \& deep learning for recommender systems},
  author={Cheng, Heng-Tze and Koc, Levent and Harmsen, Jeremiah and Shaked, Tal and Chandra, Tushar and Aradhye, Hrishi and Anderson, Glen and Corrado, Greg and Chai, Wei and Ispir, Mustafa and others},
  booktitle={Proceedings of the 1st workshop on deep learning for recommender systems},
  pages={7--10},
  year={2016}
}

@inproceedings{guo2021dual,
  title={Dual Graph enhanced Embedding Neural Network for CTR Prediction},
  author={Guo, Wei and Su, Rong and Tan, Renhao and Guo, Huifeng and Zhang, Yingxue and Liu, Zhirong and Tang, Ruiming and He, Xiuqiang},
  booktitle={Proceedings of the 27th ACM SIGKDD Conference on Knowledge Discovery \& Data Mining},
  pages={496--504},
  year={2021}
}

@inproceedings{li2019fi,
  title={Fi-gnn: Modeling feature interactions via graph neural networks for ctr prediction},
  author={Li, Zekun and Cui, Zeyu and Wu, Shu and Zhang, Xiaoyu and Wang, Liang},
  booktitle={Proceedings of the 28th ACM International Conference on Information and Knowledge Management},
  pages={539--548},
  year={2019}
}

@inproceedings{zhu2020ensembled,
  title={Ensembled CTR prediction via knowledge distillation},
  author={Zhu, Jieming and Liu, Jinyang and Li, Weiqi and Lai, Jincai and He, Xiuqiang and Chen, Liang and Zheng, Zibin},
  booktitle={Proceedings of the 29th ACM International Conference on Information \& Knowledge Management},
  pages={2941--2958},
  year={2020}
}

@inproceedings{yu2020deep,
  title={Deep interaction machine: A simple but effective model for high-order feature interactions},
  author={Yu, Feng and Liu, Zhaocheng and Liu, Qiang and Zhang, Haoli and Wu, Shu and Wang, Liang},
  booktitle={Proceedings of the 29th ACM International Conference on Information \& Knowledge Management},
  pages={2285--2288},
  year={2020}
}

@inproceedings{zhou2018rocket,
  title={Rocket launching: A universal and efficient framework for training well-performing light net},
  author={Zhou, Guorui and Fan, Ying and Cui, Runpeng and Bian, Weijie and Zhu, Xiaoqiang and Gai, Kun},
  booktitle={Proceedings of the AAAI Conference on Artificial Intelligence},
  volume={32},
  number={1},
  year={2018}
}

@inproceedings{xu2020privileged,
  title={Privileged features distillation at Taobao recommendations},
  author={Xu, Chen and Li, Quan and Ge, Junfeng and Gao, Jinyang and Yang, Xiaoyong and Pei, Changhua and Sun, Fei and Wu, Jian and Sun, Hanxiao and Ou, Wenwu},
  booktitle={Proceedings of the 26th ACM SIGKDD International Conference on Knowledge Discovery \& Data Mining},
  pages={2590--2598},
  year={2020}
}

@inproceedings{liu2019feature,
  title={Feature generation by convolutional neural network for click-through rate prediction},
  author={Liu, Bin and Tang, Ruiming and Chen, Yingzhi and Yu, Jinkai and Guo, Huifeng and Zhang, Yuzhou},
  booktitle={The World Wide Web Conference},
  pages={1119--1129},
  year={2019}
}

@article{zhu2021aim,
  title={AIM: Automatic Interaction Machine for Click-Through Rate Prediction},
  author={Zhu, Chenxu and Chen, Bo and Zhang, Weinan and Lai, Jincai and Tang, Ruiming and He, Xiuqiang and Li, Zhenguo and Yu, Yong},
  journal={IEEE Transactions on Knowledge and Data Engineering},
  year={2021},
  publisher={IEEE}
}

@inproceedings{khawar2020autofeature,
  title={Autofeature: Searching for feature interactions and their architectures for click-through rate prediction},
  author={Khawar, Farhan and Hang, Xu and Tang, Ruiming and Liu, Bin and Li, Zhenguo and He, Xiuqiang},
  booktitle={Proceedings of the 29th ACM International Conference on Information \& Knowledge Management},
  pages={625--634},
  year={2020}
}

@article{li2021graphfm,
  title={GraphFM: Graph factorization machines for feature interaction modeling},
  author={Li, Zekun and Wu, Shu and Cui, Zeyu and Zhang, Xiaoyu},
  journal={arXiv e-prints},
  pages={arXiv--2105},
  year={2021}
}

@article{zhao2020amer,
  title={Amer: Automatic behavior modeling and interaction exploration in recommender system},
  author={Zhao, Pengyu and Xiao, Kecheng and Zhang, Yuanxing and Bian, Kaigui and Yan, Wei},
  journal={arXiv preprint arXiv:2006.05933},
  year={2020}
}

@inproceedings{cheng2020adaptive,
  title={Adaptive factorization network: Learning adaptive-order feature interactions},
  author={Cheng, Weiyu and Shen, Yanyan and Huang, Linpeng},
  booktitle={Proceedings of the AAAI Conference on Artificial Intelligence},
  volume={34},
  number={04},
  pages={3609--3616},
  year={2020}
}

@article{cybenko1989approximation,
  title={Approximation by superpositions of a sigmoidal function},
  author={Cybenko, George},
  journal={Mathematics of control, signals and systems},
  volume={2},
  number={4},
  pages={303--314},
  year={1989},
  publisher={Springer}
}

@article{zou2018stochastic,
  title={Stochastic gradient descent optimizes over-parameterized deep ReLU networks},
  author={Zou, Difan and Cao, Yuan and Zhou, Dongruo and Gu, Quanquan},
  journal={arXiv preprint arXiv:1811.08888},
  year={2018}
}

@inproceedings{tian2024rotative,
  title={Rotative Factorization Machines},
  author={Tian, Zhen and Shi, Yuhong and Wu, Xiangkun and Zhao, Wayne Xin and Wen, Ji-Rong},
  booktitle={Proceedings of the 30th ACM SIGKDD Conference on Knowledge Discovery and Data Mining},
  pages={2912--2923},
  year={2024}
}

@inproceedings{lu2021dual,
  title={A dual input-aware factorization machine for CTR prediction},
  author={Lu, Wantong and Yu, Yantao and Chang, Yongzhe and Wang, Zhen and Li, Chenhui and Yuan, Bo},
  booktitle={Proceedings of the Twenty-Ninth International Conference on International Joint Conferences on Artificial Intelligence},
  pages={3139--3145},
  year={2021}
}

@article{lobo2008auc,
  title={AUC: a misleading measure of the performance of predictive distribution models},
  author={Lobo, Jorge M and Jim{\'e}nez-Valverde, Alberto and Real, Raimundo},
  journal={Global ecology and Biogeography},
  volume={17},
  number={2},
  pages={145--151},
  year={2008},
  publisher={Wiley Online Library}
}

@article{buja2005loss,
  title={Loss functions for binary class probability estimation and classification: Structure and applications},
  author={Buja, Andreas and Stuetzle, Werner and Shen, Yi},
  journal={Working draft, November},
  volume={3},
  pages={13},
  year={2005}
}

@inproceedings{tian2023eulernet,
  title = {EulerNet: Adaptive Feature Interaction Learning via Euler's Formula for CTR Prediction},
  author = {Tian, Zhen and Bai, Ting and Zhao, Wayne Xin and Wen, Ji-Rong and Cao, Zhao},
  booktitle = {Proceedings of the 46th International ACM SIGIR Conference on Research and Development in Information Retrieval},
  pages = {1376–1385},
  year = {2023},
}

@inproceedings{cai2021arm,
  title={Arm-net: Adaptive relation modeling network for structured data},
  author={Cai, Shaofeng and Zheng, Kaiping and Chen, Gang and Jagadish, HV and Ooi, Beng Chin and Zhang, Meihui},
  booktitle={Proceedings of the 2021 International Conference on Management of Data},
  pages={207--220},
  year={2021}
}

@inproceedings{wang2022enhancing,
  title={Enhancing CTR prediction with context-aware feature representation learning},
  author={Wang, Fangye and Wang, Yingxu and Li, Dongsheng and Gu, Hansu and Lu, Tun and Zhang, Peng and Gu, Ning},
  booktitle={Proceedings of the 45th International ACM SIGIR Conference on Research and Development in Information Retrieval},
  pages={343--352},
  year={2022}
}

@inproceedings{zhou2019deep,
  title={Deep interest evolution network for click-through rate prediction},
  author={Zhou, Guorui and Mou, Na and Fan, Ying and Pi, Qi and Bian, Weijie and Zhou, Chang and Zhu, Xiaoqiang and Gai, Kun},
  booktitle={Proceedings of the AAAI conference on artificial intelligence},
  volume={33},
  number={01},
  pages={5941--5948},
  year={2019}
}

@inproceedings{xiao2017attentional,
  title={Attentional factorization machines: learning the weight of feature interactions via attention networks},
  author={Xiao, Jun and Ye, Hao and He, Xiangnan and Zhang, Hanwang and Wu, Fei and Chua, Tat-Seng},
  booktitle={Proceedings of the 26th International Joint Conference on Artificial Intelligence},
  pages={3119--3125},
  year={2017}
}

@inproceedings{qu2017product,
  title={Product-Based Neural Networks for User Response Prediction},
  author={Qu, Y and Cai, H and Zhang, W and Wen, Y and Wang, J},
  booktitle={The IEEE International Conference on Data Mining},
  pages={1149--1154},
  year={2017},
  organization={IEEE}
}

@inproceedings{zhu2023final,
  title={Final: Factorized interaction layer for ctr prediction},
  author={Zhu, Jieming and Jia, Qinglin and Cai, Guohao and Dai, Quanyu and Li, Jingjie and Dong, Zhenhua and Tang, Ruiming and Zhang, Rui},
  booktitle={Proceedings of the 46th International ACM SIGIR Conference on Research and Development in Information Retrieval},
  pages={2006--2010},
  year={2023}
}

@inproceedings{feng2019deep,
  title={Deep session interest network for click-through rate prediction},
  author={Feng, Yufei and Lv, Fuyu and Shen, Weichen and Wang, Menghan and Sun, Fei and Zhu, Yu and Yang, Keping},
  booktitle={Proceedings of the 28th International Joint Conference on Artificial Intelligence},
  pages={2301--2307},
  year={2019}
}

@inproceedings{mao2023finalmlp,
  title={FinalMLP: an enhanced two-stream MLP model for CTR prediction},
  author={Mao, Kelong and Zhu, Jieming and Su, Liangcai and Cai, Guohao and Li, Yuru and Dong, Zhenhua},
  booktitle={Proceedings of the AAAI Conference on Artificial Intelligence},
  volume={37},
  number={4},
  pages={4552--4560},
  year={2023}
}

@inproceedings{wang2023towards,
  title={Towards deeper, lighter and interpretable cross network for ctr prediction},
  author={Wang, Fangye and Gu, Hansu and Li, Dongsheng and Lu, Tun and Zhang, Peng and Gu, Ning},
  booktitle={Proceedings of the 32nd ACM International Conference on Information and Knowledge Management},
  pages={2523--2533},
  year={2023}
}

@article{wang2021masknet,
  title={Masknet: Introducing feature-wise multiplication to CTR ranking models by instance-guided mask},
  author={Wang, Zhiqiang and She, Qingyun and Zhang, Junlin},
  journal={arXiv preprint arXiv:2102.07619},
  year={2021}
}

\end{document}